\documentclass[onecolumn,showpacs,preprintnumbers]{revtex4}
\usepackage{mathrsfs}
\usepackage{graphicx}
\usepackage{dcolumn}
\usepackage{bm}
\usepackage{epsfig}
\usepackage{amsmath,amssymb}

\begin{document}
\title{The $1D$, $2D$ $\Xi_{b}$ and $\Lambda_{b}$ baryons}
\author{Guo-Liang Yu$^{1}$}
\email{yuguoliang2011@163.com}
\author{Zhi-Gang Wang$^{1}$}
\email{zgwang@aliyun.com}
\author{Xiu-Wu Wang}

\affiliation{Department of Mathematics and Physics, North China
Electric power university, Baoding 071003, People's Republic of
China}
\date{\today }

\begin{abstract}
Recently, scientists have made great progresses in experiments in searching for the excited $\Xi_{b}$ and $\Lambda_{b}$ baryons such as the $\Lambda_{b}(6072)$, $\Lambda_{b}(6146)$, $\Lambda_{b}(6152)$, $\Xi_{b}(6227)$, $\Xi_{b}(6100)$, $\Xi_{b}(6327)$ and $\Xi_{b}(6333)$. Motivated by these progresses, we give a systematical analysis about the $1D$ and $2D$ states of $\Xi_{b}$ and $\Lambda_{b}$ baryons with the method of QCD sum rules. By constructing three types of interpolating currents, we calculate the masses and pole residues of these heavy baryons with different excitation modes $(L_{\rho},L_{\lambda})=(0,2)$, $(2,0)$ and $(1,1)$.
As a result, we decode the inner structure of $\Lambda_{b}(6146)$, $\Lambda_{b}(6152)$, $\Xi_{b}(6327)$ and $\Xi_{b}(6333)$, and favor assigning these states as the $1D$ baryons with the quantum numbers $(L_{\rho},L_{\lambda})=(0,2)$ and $\frac{3}{2}^{+}$, $\frac{5}{2}^{+}$, $\frac{3}{2}^{+}$ and $\frac{5}{2}^{+}$, respectively.  In addition, the predictions for the masses and pole residues of the other $1D$, $2D$ $\Xi_{b}$ and $\Lambda_{b}$ baryons in this paper are helpful in studying the D-wave bottom baryons in the future.
\end{abstract}


\maketitle

\begin{large}
\textbf{1 Introduction}
\end{large}

In recent years, more and more heavy baryons have been confirmed by Belle, LHCb and CDF collaborations and the spectra of the charm and bottom baryon families have become more and more abundance. Especially for the excited bottom baryons, scientists have made great progresses in theoretical and experimental studies in recent years, such as the $\Lambda_{b}(5912)$ and $\Lambda_{b}(5920)$\cite{LHCb1,CDF1}, $\Lambda_{b}(6072)$\cite{60720,60721}, $\Xi_{b}(6227)$\cite{Xi62270,Xi62271,Xi62272}, $\Xi_{b}(6100)$\cite{Xi61000}. In 2019, the LHCb Collaboration reported the discovery of two bottom baryon states, $\Lambda_{b}(6146)^{0}$ and  $\Lambda_{b}(6152)^{0}$, by analyzing the $\Lambda_{b}^{0}\pi^{+}\pi^{-}$ invariant mass spectrum from $pp$ collisions\cite{LHCb2}. The measured masses and widths are
\begin{center}
$m_{\Lambda_{b}(6146)^{0}}=6146.17\pm0.33\pm0.22\pm0.16$ MeV, $\Gamma_{\Lambda_{b}(6146)^{0}}=2.9\pm1.3\pm0.3$ MeV,

$m_{\Lambda_{b}(6152)^{0}}=6152.51\pm0.26\pm0.22\pm0.16$ MeV, $\Gamma_{\Lambda_{b}(6152)^{0}}=2.1\pm2.1\pm0.8\pm0.3$ MeV.
\end{center}
By studying their strong decays using quark model or $^{3}P_{0}$ model, people interpreted these two states as a $\Lambda_{b}(1D)$ doublet\cite{L61460,L61461,L61462,L61463,Xi62271}. Before this observation, different collaborations predicted the masses of this doublet with the quark model\cite{quam3,quam6,quam9,Theo2}, whose results were not consistent well with each other and with the experiments, and need further confirmations by different theoretical metheds/models.

Very recently, LHCb collaboration reported the observation of two new excited $\Xi_{b}$ states in the $\Lambda_{b}K^{-}\pi^{+}$ mass spectrum\cite{Xi6333}. The measured masses and widths are

\begin{center}
$m_{\Xi_{b}(6327)}=6327.28^{+0.23}_{-0.21}$(stat)$\pm0.08$(syst)$\pm0.24(m_{\Lambda{b}})$ MeV $\Gamma_{\Xi_{b}(6327)}<2.20$ MeV

$m_{\Xi_{b}(6333)}=6332.69^{+0.17}_{-0.18}$(stat)$\pm0.03$(syst)$\pm0.22(m_{\Lambda{b}})$ MeV $\Gamma_{\Xi_{b}(6327)}<1.55$ MeV
\end{center}
By comparing with the quark-model predictions\cite{Xi62271,L61460}, Chen et al. interpreted these two states as a $1D$($\Xi_{b}$) doublet with $J^{P}=\frac{3}{2}^{+}$ and $\frac{5}{2}^{+}$.

People have adopted many theoretical methods/models in the past decades to investigate the bottom baryons, including quark model\cite{L61462,quam1,quam2,quam3,quam4,quam5,quam6,quam7,quam8,quam9,quam10,quam11,quam12,quam13,quam14,quam15,quam16,quam17,quam18,quam19,quam20,quam21,quam22,quam23,quam26,quam27,quam28}, heavy hadron chiral perturbation theory\cite{chiral1,chiral2,chiral3,chiral4,chiral5,chiral6}, $^{3}P_{0}$ decay model\cite{3P01,3P02,3P03,3P04,3P05,3P06,3P07}, lattice QCD\cite{Lattice1,Lattice2,Lattice3,Lattice4}, light cone QCD sum rules\cite{LCsum1,LCsum2,LCsum3,LCsum4,LCsum5,LCsum6,LCsum7,LCsum8} and QCD sum rules\cite{Sum1,Sum2,Sum3,Sum4,Sum5,Sum6,Sum7,Sum8,Sum9,Sum10}, etc. For more discussions about the bottom baryon states, one can consult the Refs.\cite{Theo1,Theo2,Theo3,Theo4,Theo5,Theo6,Theo7,Theo8,Theo9,Theo10} and the references therein. With the efforts of theoretical and experimental physicists, some bottom baryon states have been observed and confirmed such as $\Xi_{b}(5797)$\cite{Xi5797}, $\Lambda_{b}(5620)$\cite{Xi5797}, $\Lambda_{b}(5912)$\cite{Xi5797}, $\Lambda_{b}(5920)$\cite{Xi5797}, $\Lambda_{b}(6072)$\cite{60720,60721} whose quantum numbers were determined to be 1S($\frac{1}{2}^{+}$), 1S($\frac{1}{2}^{+}$), 1P($\frac{1}{2}^{-}$), 1P($\frac{3}{2}^{-}$) and 2S($\frac{1}{2}^{+}$), respectively. However, the inner structure of the newly observed baryon states $\Xi_{b}(6327)$, $\Xi_{b}(6333)$, $\Lambda(6146)$ and $\Lambda(6152)$ needs further confirmation theoretically. The other bottom baryon states, such as the radially excited D-wave $\Xi_{b}$ and $\Lambda_{b}$ baryons, have not been observed.

QCD sum rules has been proved to be a most powerful non-perturbative method in studying the properties of the mesons and baryons\cite{sumrule1,sumrule2,Sum4,baryon1,baryon2,baryon3,baryon4,baryon5,baryon6,gly}, and it has been extended to study the multiquark states\cite{sumrule3,sumrule4,sumrule5,sumrule6,sumrule7,sumrule8,sumrule9,sumrule10}.
In our previous work, we systematically studied the D-wave charmed baryons $\Lambda_{c}(2860)$, $\Lambda_{c}(2880)$, $\Xi_{c}(3055)$, $\Xi_{c}(3080)$\cite{WZG0}, the P-wave $\Omega_{c}$ states, $\Omega_{c}(3000)$, $\Omega_{c}(3050)$, $\Omega_{c}(3066)$, $\Omega_{c}(3090)$, $\Omega_{c}(3119)$\cite{WZG1,WZG2}, and the $\Omega_{b}$ states, $\Omega_{b}(6316)$, $\Omega_{b}(6330)$, $\Omega_{b}(6340)$, $\Omega_{b}(6350)$\cite{WZG3}, using the method of QCD sum rules. As a continuation of our previous work, we study the $1D$ and $2D$ states of $\Xi_{b}$ and $\Lambda_{b}$ baryons with orbital excitations ($L_{\rho},L_{\lambda})=$($0,2$), ($2,0$) and ($1,1$). The motivation of this work is to further confirm the structure of $\Lambda_{b}(6146)$, $\Lambda_{b}(6152)$, $\Xi_{b}(6327)$ and $\Xi_{b}(6333)$, decode their excitation modes and predict the masses and pole residues of the $1D$, $2D$ $\Xi_{b}$ and $\Lambda_{b}$ baryons.

The layout of this paper is as follows: in Sec.2,we first construct three types of interpolating currents for D-wave bottom baryons $\Lambda_{b}$ and $\Xi_{b}$; in Sec.3 we derive QCD sum rules for the masses and
pole residues of these states with spin-parity $\frac{3}{2}^{+}$ and $\frac{5}{2}^{+}$ from two-point correlation function; in Sect.4, we present the numerical results and discussions; and Sec.5 is reserved for our conclusions.

\begin{large}
\textbf{2 Interpolating currents for the D-wave bottom baryons}
\end{large}

In the heavy quark limit, one heavy quark within a heavy baryon system is decoupled from two light quarks. Under this scenario, the dynamics of a heavy baryon state can be well separated into two parts, the $\rho-$mode which is for the degree of freedom between two light quarks, and the $\lambda-$mode which denotes the degree between the center of mass of  diquarks and the heavy quark. In this diquark-quark model, the orbital angular momentum between the two light quarks is denoted by $L_{\rho}$, while the angular momentum between the light diquarks and the heavy quark is denoted by $L_{\lambda}$. For D-wave($L=2$) bottom baryon, there are three orbital excitation modes ($L_{\rho},L_{\lambda}$)=($2,0$), ($0,2$) and ($1,1$). The color antitriplet diquarks with quantum numbers of $L_{\rho}=0$ and $s_{l}=0$ can be written as $\varepsilon^{ijk}q^{T}_{j}C\gamma_{5}q'_{k}$ which has the  spin-parity of $J^{P}_{d}=0^{+}_{d}$. The spin-parity of relative P-wave and D-wave is denoted as $J_{\rho/\lambda}^{P}=L_{\rho/\lambda}^{P}=1_{\rho/\lambda}^{-}$ and $2_{\rho/\lambda}^{+}$, respectively. If $J_{b}^{P}=\frac{1}{2}_{b}^{+}$ is denoted as the spin-parity of b-quark, we can obtain the final states of D-wave bottom baryons according to direct product of angular momentum $J^{P}=0^{+}_{d}\bigotimes J_{\rho/\lambda}^{P}\bigotimes \frac{1}{2}_{b}^{+}$.

For the excitation mode ($L_{\rho},L_{\lambda}$)=($1,0$), the P-wave diquark system with $J^{P}=1^{-}$ can be constructed by applying derivative between two light quarks,
\begin{eqnarray}
\epsilon^{ijk}\big[\partial^{\beta}q_{i}^{T}(x)C\gamma_{5}q'_{j}(x)-q_{i}^{T}(x)C\gamma_{5}\partial^{\beta}q'_{j}(x)\big]
\end{eqnarray}
On this basis, we continue to introduce an additional derivatives between the two light quarks in Eq.(1) to obtain excitation mode of ($L_{\rho},L_{\lambda}$)=($2,0$)
\begin{eqnarray}
\notag
&&\epsilon^{ijk}\Big\{\big[\partial^{\alpha}\partial^{\beta}q_{i}^{T}(x)C\gamma_{5}q'_{j}(x)-\partial^{\beta}q_{i}^{T}(x)C\gamma_{5}\partial^{\alpha}q'_{j}(x)\big]-\big[\partial^{\alpha}q_{i}^{T}(x)C\gamma_{5}\partial^{\beta}q'_{j}(x)\\
&&-q_{i}^{T}(x)C\gamma_{5}\partial^{\alpha}\partial^{\beta}q'_{j}(x)\big]\Big\}
\end{eqnarray}
For the excitation mode ($L_{\rho},L_{\lambda}$)=($0,2$), we need apply two derivatives between the diquark system and the b-quark field. It should be noticed that the b-quark in the bottom baryon is static in the heavy quark limit. Then, the $\overleftrightarrow{\partial}_{\mu}$ is reduced to $\overleftarrow{\partial}_{\mu}$ when operating on the b-quark field and the light diquark state with $J^{P}=2^{+}$ is written as,
\begin{eqnarray}
\notag
\partial^{\alpha}\partial^{\beta}\big[\epsilon^{ijk}q_{i}^{T}(x)C\gamma_{5}q'_{j}(x)\big]=&&\epsilon^{ijk}\big[\partial^{\alpha}\partial^{\beta}q_{i}^{T}(x)C\gamma_{5}q'_{j}(x)+\partial^{\beta}q_{i}^{T}(x)C\gamma_{5}\partial^{\alpha}q'_{j}(x)\big] \\ && +\partial^{\alpha}q_{i}^{T}(x)C\gamma_{5}\partial^{\beta}q'_{j}(x)+q_{i}^{T}(x)C\gamma_{5}\partial^{\alpha}\partial^{\beta}q'_{j}(x)\big]
\end{eqnarray}
For ($L_{\rho},L_{\lambda}$)=($1,1$) state, we need an additional derivatives between the P-wave diquark(Eq.(1)) and the b-quark field,
\begin{eqnarray}
\notag
&&\partial^{\alpha}\epsilon^{ijk}\big[\partial^{\beta}q_{i}^{T}(x)C\gamma_{5}q'_{j}(x)-q_{i}^{T}(x)C\gamma_{5}\partial^{\beta}q'_{j}(x)\big]\\
&& \notag= \epsilon^{ijk}\big[\partial^{\alpha}\partial^{\beta} q_{i}^{T}(x)C\gamma_{5}q'_{j}(x)+\partial^{\beta}q_{i}^{T}(x)C\gamma_{5}\partial^{\alpha}q'_{j}(x)-\partial^{\alpha}q_{i}^{T}(x)C\gamma_{5}\partial^{\beta}q'_{j}(x) \\ && - q_{i}^{T}(x)C\gamma_{5}\partial^{\alpha}\partial^{\beta} q'_{j}(x)\big]\Gamma_{\alpha\beta\mu\nu}c_{k}(x)
\end{eqnarray}
Considering the symmetrization of the Lorentz indexes $\mu$ and $\nu$, the light diquark state with ($L_{\rho},L_{\lambda}$)=($1,1$) can be expressed in a more simple form,
\begin{eqnarray}
\epsilon^{ijk}\big[\partial^{\alpha}\partial^{\beta} q_{i}^{T}(x)C\gamma_{5}q'_{j}(x)- q_{i}^{T}(x)C\gamma_{5}\partial^{\alpha}\partial^{\beta} q'_{j}(x)\big]
\end{eqnarray}
Finally, we combine these above light diquark systems with the b-quark field to form $J^{P}=\frac{3}{2}^{+}$ or $\frac{5}{2}^{+}$ baryon states which have three excitation modes ($L_{\rho},L_{\lambda}$)=($2,0$), ($0,2$) and ($1,1$).
For more details about the construction of the interpolating currents of baryons, one can consult the Refs.\cite{baryon6, current, WZG0}. We can now classify these constructed interpolating currents as follows,
\begin{center}
 $(L_{\rho},L_{\lambda})=(0,2)$ for $J^{1}_{\mu}/\eta_{\mu}^{1}(x)$, $J^{1}_{\mu\nu}/\eta_{\mu\nu}^{1}(x)$,

 $(L_{\rho},L_{\lambda})=(2,0)$ for $J^{2}_{\mu}/\eta_{\mu}^{2}(x)$, $J^{2}_{\mu\nu}/\eta_{\mu\nu}^{2}(x)$,

 $(L_{\rho},L_{\lambda})=(1,1)$ for $J^{3}_{\mu}/\eta_{\mu}^{3}(x)$, $J^{3}_{\mu\nu}/\eta_{\mu\nu}^{3}(x)$,
\end{center}
with
\begin{eqnarray}
\notag
 J_{\mu}^{1}(x)=&&\epsilon^{ijk}\big[\partial^{\alpha}\partial^{\beta} q_{i}^{T}(x)C\gamma_{5}s_{j}(x)+\partial^{\alpha}q_{i}^{T}(x)C\gamma_{5}\partial^{\beta}s_{j}(x)+\partial^{\beta}q_{i}^{T}(x)C\gamma_{5}\partial^{\alpha}s_{j}(x) \\&&
 \notag+ q_{i}^{T}(x)C\gamma_{5}\partial^{\alpha}\partial^{\beta} s_{j}(x)\big]\Gamma_{\alpha\beta\mu}b_{k}(x)\\
\notag
J_{\mu}^{2}(x)=&&\epsilon^{ijk}\big[\partial^{\alpha}\partial^{\beta} q_{i}^{T}(x)C\gamma_{5}s_{j}(x)-\partial^{\alpha}q_{i}^{T}(x)C\gamma_{5}\partial^{\beta}s_{j}(x)-\partial^{\beta}q_{i}^{T}(x)C\gamma_{5}\partial^{\alpha}s_{j}(x)
\\ && \notag+ q_{i}^{T}(x)C\gamma_{5}\partial^{\alpha}\partial^{\beta} s_{j}(x)\big]\Gamma_{\alpha\beta\mu}b_{k}(x) \\
J_{\mu}^{3}(x)=&&\epsilon^{ijk}\big[\partial^{\alpha}\partial^{\beta} q_{i}^{T}(x)C\gamma_{5}s_{j}(x)- q_{i}^{T}(x)C\gamma_{5}\partial^{\alpha}\partial^{\beta} s_{j}(x)\big]\Gamma_{\alpha\beta\mu}b_{k}(x)
\end{eqnarray}
\begin{eqnarray}
\notag
\eta_{\mu}^{1}(x)=&&\epsilon^{ijk}\big[\partial^{\alpha}\partial^{\beta} q_{i}^{T}(x)C\gamma_{5}q'_{j}(x)+\partial^{\alpha}q_{i}^{T}(x)C\gamma_{5}\partial^{\beta}q'_{j}(x)+\partial^{\beta}q_{i}^{T}(x)C\gamma_{5}\partial^{\alpha}q'_{j}(x) \\ && \notag
+ q_{i}^{T}(x)C\gamma_{5}\partial^{\alpha}\partial^{\beta} q'_{j}(x)\big]\Gamma_{\alpha\beta\mu}b_{k}(x) \\
\notag
\eta_{\mu}^{2}(x)=&&\epsilon^{ijk}\big[\partial^{\alpha}\partial^{\beta} q_{i}^{T}(x)C\gamma_{5}q'_{j}(x)-\partial^{\alpha}q_{i}^{T}(x)C\gamma_{5}\partial^{\beta}q'_{j}(x)-\partial^{\beta}q_{i}^{T}(x)C\gamma_{5}\partial^{\alpha}q'_{j}(x) \\ && \notag + q_{i}^{T}(x)C\gamma_{5}\partial^{\alpha}\partial^{\beta} q'_{j}(x)\big]\Gamma_{\alpha\beta\mu}b_{k}(x) \\
\eta_{\mu}^{3}(x)=&&\epsilon^{ijk}\big[\partial^{\alpha}\partial^{\beta} q_{i}^{T}(x)C\gamma_{5}q'_{j}(x)- q_{i}^{T}(x)C\gamma_{5}\partial^{\alpha}\partial^{\beta} q'_{j}(x)\big]\Gamma_{\alpha\beta\mu}b_{k}(x)
\end{eqnarray}
\begin{eqnarray}
\notag
J_{\mu\nu}^{1}(x)=&&\epsilon^{ijk}\big[\partial^{\alpha}\partial^{\beta} q_{i}^{T}(x)C\gamma_{5}s_{j}(x)+\partial^{\alpha}q_{i}^{T}(x)C\gamma_{5}\partial^{\beta}s_{j}(x)+\partial^{\beta}q_{i}^{T}(x)C\gamma_{5}\partial^{\alpha}s_{j}(x) \\ && \notag + q_{i}^{T}(x)C\gamma_{5}\partial^{\alpha}\partial^{\beta} s_{j}(x)\big]\Gamma_{\alpha\beta\mu\nu}b_{k}(x) \\
\notag
J_{\mu\nu}^{2}(x)=&&\epsilon^{ijk}\big[\partial^{\alpha}\partial^{\beta} q_{i}^{T}(x)C\gamma_{5}s_{j}(x)-\partial^{\alpha}q_{i}^{T}(x)C\gamma_{5}\partial^{\beta}s_{j}(x)-\partial^{\beta}q_{i}^{T}(x)C\gamma_{5}\partial^{\alpha}s_{j}(x) \\ && \notag+ q_{i}^{T}(x)C\gamma_{5}\partial^{\alpha}\partial^{\beta} s_{j}(x)\big]\Gamma_{\alpha\beta\mu\nu}b_{k}(x) \\
J_{\mu\nu}^{3}(x)=&&\epsilon^{ijk}\big[\partial^{\alpha}\partial^{\beta} q_{i}^{T}(x)C\gamma_{5}s_{j}(x)- q_{i}^{T}(x)C\gamma_{5}\partial^{\alpha}\partial^{\beta} s_{j}(x)\big]\Gamma_{\alpha\beta\mu\nu}b_{k}(x)
\end{eqnarray}
\begin{eqnarray}
\notag
\eta_{\mu\nu}^{1}(x)=&&\epsilon^{ijk}\big[\partial^{\alpha}\partial^{\beta} q_{i}^{T}(x)C\gamma_{5}q'_{j}(x)+\partial^{\alpha}q_{i}^{T}(x)C\gamma_{5}\partial^{\beta}q'_{j}(x)+\partial^{\beta}q_{i}^{T}(x)C\gamma_{5}\partial^{\alpha}q'_{j}(x) \\ && \notag+ q_{i}^{T}(x)C\gamma_{5}\partial^{\alpha}\partial^{\beta} q'_{j}(x)\big]\Gamma_{\alpha\beta\mu\nu}b_{k}(x) \\
\notag
\eta_{\mu\nu}^{2}(x)=&&\epsilon^{ijk}\big[\partial^{\alpha}\partial^{\beta} q_{i}^{T}(x)C\gamma_{5}q'_{j}(x)-\partial^{\alpha}q_{i}^{T}(x)C\gamma_{5}\partial^{\beta}q'_{j}(x)-\partial^{\beta}q_{i}^{T}(x)C\gamma_{5}\partial^{\alpha}q'_{j}(x) \\ && \notag+ q_{i}^{T}(x)C\gamma_{5}\partial^{\alpha}\partial^{\beta} q'_{j}(x)\big]\Gamma_{\alpha\beta\mu\nu}b_{k}(x) \\
\eta_{\mu\nu}^{3}(x)=&&\epsilon^{ijk}\big[\partial^{\alpha}\partial^{\beta} q_{i}^{T}(x)C\gamma_{5}q'_{j}(x)- q_{i}^{T}(x)C\gamma_{5}\partial^{\alpha}\partial^{\beta} q'_{j}(x)\big]\Gamma_{\alpha\beta\mu\nu}b_{k}(x)
\end{eqnarray}

where $\Gamma_{\alpha\beta\mu}$ and $\Gamma_{\alpha\beta\mu\nu}$ are the projection operatiors, whose explicit form is,
\begin{eqnarray}
\Gamma_{\alpha\beta\mu}=(g_{\alpha\mu}g_{\beta\nu}+g_{\alpha\nu}g_{\beta\mu}-\frac{1}{2}g_{\alpha\beta}g_{\mu\nu})\gamma^{\nu}\gamma_{5}
\end{eqnarray}
\begin{eqnarray}
\notag
\Gamma_{\alpha\beta\mu\nu}=&&g_{\alpha\mu}g_{\beta\nu}+g_{\alpha\nu}g_{\beta\mu}-\frac{1}{6}g_{\alpha\beta}g_{\mu\nu}-\frac{1}{4}g_{\alpha\mu}\gamma_{\beta}\gamma_{\nu}-\frac{1}{4}g_{\alpha\nu}\gamma_{\beta}\gamma_{\mu}-\frac{1}{4}g_{\beta\mu}\gamma_{\alpha}\gamma_{\nu}-\frac{1}{4}g_{\beta\nu}\gamma_{\alpha}\gamma_{\mu} \\ && +\frac{1}{24}\gamma_{\alpha}\gamma_{\mu}\gamma_{\beta}\gamma_{\nu}+\frac{1}{24}\gamma_{\alpha}\gamma_{\nu}\gamma_{\beta}\gamma_{\mu}+\frac{1}{24}\gamma_{\beta}\gamma_{\mu}\gamma_{\alpha}\gamma_{\nu}+\frac{1}{24}\gamma_{\beta}\gamma_{\nu}\gamma_{\alpha}\gamma_{\mu}
\end{eqnarray}

\begin{large}
\textbf{3  QCD sum rules for the $1D$ and $2D$ $\Xi_{b}$ and $\Lambda_{b}$ states}
\end{large}

The first step of the analysis with QCD sum rules is to write down the following
two-point correlation functions,
\begin{eqnarray}
\notag
\Pi_{\mu\nu}(p)&&=i\int d^{4}xe^{ip.x}\Big\langle0|T\Big\{J_{\mu}/\eta_{\mu}(x)\overline{J}_{\nu}/\overline{\eta}_{\nu}(0)\Big\}
|0\Big\rangle\\
\Pi_{\mu\nu\alpha\beta}(p)&&=i\int d^{4}xe^{ip.x}\Big\langle0|T\Big\{J_{\mu\nu}/\eta_{\mu\nu}(x)\overline{J}_{\alpha\beta}/\overline{\eta}_{\alpha\beta}(0)\Big\}
|0\Big\rangle
\end{eqnarray}
where $\emph{T}$ is the time ordered product. The currents $J_{\mu}/\eta_{\mu}(0)$ and $J_{\mu\nu}/\eta_{\mu\nu}(0)$ in these above correlations couple potentially to $1D$ bottom states $B_{\frac{3}{2}^{\pm}}$ and $B_{\frac{5}{2}^{\pm}}$, respectively and couple also to $2D$ states $B^{\prime}_{\frac{3}{2}^{\pm}}$ and $B^{\prime}_{\frac{5}{2}^{\pm}}$ with the quantum numbers $\frac{3}{2}^{+}$ and $\frac{5}{2}^{+}$,
\begin{eqnarray}
\notag
&&\langle 0| J/\eta_{\mu}(0)|B^{(\prime)+}_{\frac{3}{2}}(p) \rangle=\lambda^{(\prime)+}_{\frac{3}{2}}U^{+}_{\mu}(p,s),  \\
&&\langle 0| J/\eta_{\mu\nu}(0)|B^{(\prime)+}_{\frac{5}{2}}(p) \rangle=\lambda^{(\prime)+}_{\frac{5}{2}}U^{+}_{\mu\nu}(p,s), \\
\notag
&&\langle 0| J/\eta_{\mu}(0)|B^{(\prime)-}_{\frac{3}{2}}(p) \rangle=\lambda^{(\prime)-}_{\frac{3}{2}}i\gamma_{5}U^{-}_{\mu}(p,s),\\
&&\langle 0| J/\eta_{\mu\nu}(0)|B^{(\prime)-}_{\frac{5}{2}}(p) \rangle=\lambda^{(\prime)-}_{\frac{5}{2}}i\gamma_{5}U^{-}_{\mu\nu}(p,s)
\end{eqnarray}

\begin{large}
\textbf{3.1 The Phenomenological side}
\end{large}

At the hadron level, a complete set of intermediate baryon states with the same quantum
numbers as the current operators $J_{\mu}/\eta_{\mu}(x)$, $J_{\mu\nu}/\eta_{\mu\nu}(x)$, $i\gamma_{5}J_{\mu}/\eta_{\mu}(x)$ and $i\gamma_{5}J_{\mu\nu}/\eta_{\mu\nu}(x)$ are inserted
into the correlation functions $\Pi_{\mu\nu}(p)$ and $\Pi_{\mu\nu\alpha\beta}(p)$. After separating the pole terms of the lowest $1D$ and $2D$ states, we obtain the following results,
\begin{eqnarray}
\notag
\Pi_{\mu\nu}(p)=&&(\lambda^{+2}_{\frac{3}{2}}\frac{p\!\!\!/+M_{\frac{3}{2}}^{+}}{M_{\frac{3}{2}}^{+2}-p^{2}}+\lambda^{-2}_{\frac{3}{2}}\frac{p\!\!\!/-M_{\frac{3}{2}}^{-}}{M_{\frac{3}{2}}^{-2}-p^{2}}
+\lambda^{\prime+2}_{\frac{3}{2}}\frac{p\!\!\!/+M^{\prime+}_{\frac{3}{2}}}{M_{\frac{3}{2}}^{\prime+2}-p^{2}}+\lambda^{\prime-2}_{\frac{3}{2}}\frac{p\!\!\!/-M^{\prime-}_{\frac{3}{2}}}{M_{\frac{3}{2}}^{\prime-2}-p^{2}}) \\
\notag
&&\times(-g_{\mu\nu}+\frac{\gamma_{\mu}\gamma_{\nu}}{3}+\frac{2p_{\mu}p_{\nu}}{3p^{2}}-\frac{p_{\mu}\gamma_{\nu}-p_{\nu}\gamma_{\mu}}{3\sqrt{p^{2}}})+\cdots \\
&&=\Pi_{\frac{3}{2}}(p^{2})(-g_{\mu\nu})+\cdots
\end{eqnarray}
\begin{eqnarray}
\notag
\Pi_{\mu\nu}(p)=&&(\lambda^{+2}_{\frac{5}{2}}\frac{p\!\!\!/+M_{\frac{5}{2}}^{+}}{M_{\frac{5}{2}}^{+2}-p^{2}}+\lambda^{-2}_{\frac{5}{2}}\frac{p\!\!\!/-M_{\frac{5}{2}}^{-}}{M_{\frac{5}{2}}^{-2}-p^{2}}
+\lambda^{\prime+2}_{\frac{5}{2}}\frac{p\!\!\!/+M^{\prime+}_{\frac{5}{2}}}{M_{\frac{5}{2}}^{\prime+2}-p^{2}}+\lambda^{\prime-2}_{\frac{5}{2}}\frac{p\!\!\!/-M^{\prime-}_{\frac{5}{2}}}{M_{\frac{5}{2}}^{\prime-2}-p^{2}}) \\
\notag
&& \times\Big[\frac{\widetilde{g}_{\alpha\mu}\widetilde{g}_{\beta\nu}
+\widetilde{g}_{\alpha\nu}\widetilde{g}_{\beta\mu}}{2}-\frac{\widetilde{g}_{\alpha\beta}\widetilde{g}_{\mu\nu}}{5}-\frac{1}{10}(\gamma_{\mu}\gamma_{\alpha}+\frac{p_{\alpha}\gamma_{\mu}-p_{\mu}\gamma_{\alpha}}{\sqrt{p^{2}}}-\frac{p_{\mu}p_{\alpha}}{p^{2}}) \widetilde{g}_{\beta\nu} \\ \notag
&& -\frac{1}{10}(\gamma_{\mu}\gamma_{\beta}+\frac{p_{\beta}\gamma_{\mu}-p_{\mu}\gamma_{\beta}}{\sqrt{p^{2}}}-\frac{p_{\mu}p_{\beta}}{p^{2}})\widetilde{g}_{\alpha\mu}\Big]+\cdots\\
&&=\Pi_{\frac{5}{2}}(p^{2})\frac{g_{\alpha\mu}g_{\beta\nu}+g_{\alpha\nu}g_{\beta\mu}}{2}+\cdots,
\end{eqnarray}
where $\widetilde{g}_{\mu\nu}=g_{\mu\nu}-\frac{p_{\mu}p_{\nu}}{p^{2}}$, $M_{j}^{+}$, $M_{j}^{\prime+}$ denote the masses of $1D$ and $2D$ states with positive parity and angular momentum $j$, $M_{j}^{-}$, $M_{j}^{\prime-}$ are for the states with negative parity. In these derivations, we use the following relations about the spinors $U_{\mu}^{\pm}(p,s)$ and $U_{\mu\nu}^{\pm}(p,s)$,
\begin{eqnarray}
\mathop{\sum}\limits_sU_{\mu}\overline{U}_{\nu}=(p\!\!\!/+M^{(\prime)\pm})(-g_{\mu\nu}+\frac{\gamma_{\mu}\gamma_{\nu}}{3}+\frac{2p_{\mu}p_{\nu}}{3p^{2}}-\frac{p_{\mu}\gamma_{\nu}-p_{\nu}\gamma_{\mu}}{3\sqrt{p^{2}}})
\end{eqnarray}
\begin{eqnarray}
\notag
\mathop{\sum}\limits_sU_{\alpha\beta}\overline{U}_{\mu\nu}=&&(p\!\!\!/+M^{(\prime)\pm})\Big\{\frac{\widetilde{g}_{\alpha\mu}\widetilde{g}_{\beta\nu}
+\widetilde{g}_{\alpha\nu}\widetilde{g}_{\beta\mu}}{2}-\frac{\widetilde{g}_{\alpha\beta}\widetilde{g}_{\mu\nu}}{5}
-\frac{1}{10}(\gamma_{\alpha}\gamma_{\mu}+\frac{p_{\mu}\gamma_{\alpha}-p_{\alpha}\gamma_{\mu}}{\sqrt{p^{2}}}-\frac{p_{\alpha}p_{\mu}}{p^{2}})\widetilde{g}_{\beta\nu}\\
 \notag && -\frac{1}{10}(\gamma_{\beta}\gamma_{\mu}+\frac{p_{\mu}\gamma_{\beta}-p_{\beta}\gamma_{\mu}}{\sqrt{p^{2}}}-\frac{p_{\beta}p_{\mu}}{p^{2}})\widetilde{g}_{\alpha\nu}
  -\frac{1}{10}(\gamma_{\alpha}\gamma_{\nu}+\frac{p_{\nu}\gamma_{\alpha}-p_{\alpha}\gamma_{\nu}}{\sqrt{p^{2}}}-\frac{p_{\alpha}p_{\nu}}{p^{2}})\widetilde{g}_{\beta\mu} \\ &&   -\frac{1}{10}(\gamma_{\beta}\gamma_{\nu}+\frac{p_{\nu}\gamma_{\beta}-p_{\beta}\gamma_{\nu}}{\sqrt{p^{2}}}-\frac{p_{\beta}p_{\nu}}{p^{2}})\widetilde{g}_{\alpha\mu}\Big\},
\end{eqnarray}
with $p^{2}=M^{\pm2}$ and $p^{2}=M^{(\prime)\pm2}$ on mass-shell for $1D$ and $2D$ states, respectively.
From the imaginary part, we can obtain the spectral densities at the hadron side,
\begin{eqnarray}
\notag
\frac{Im\Pi_{j}(s)}{\pi}&&=p\!\!\!/\big[\lambda^{+2}_{j}\delta(s-M_{j}^{+2})+\lambda^{-2}_{j}\delta(s-M_{j}^{-2})+\lambda^{\prime+2}_{j}\delta(s-M_{j}^{\prime+2})+\lambda^{\prime-2}_{j}\delta(s-M_{j}^{\prime-2})\big]\\
\notag
&&+\big[M_{j}^{+}\lambda^{+2}_{j}\delta(s-M_{j}^{+2})-M_{j}^{-}\lambda^{-2}_{j}\delta(s-M_{j}^{-2})+M^{\prime+}_{j}\lambda^{\prime+2}_{j}\delta(s-M_{j}^{\prime+2})-M^{\prime-}_{j}\lambda^{\prime-2}_{j}\delta(s-M_{j}^{\prime-2})\big],\\
&&= p\!\!\!/\rho_{j,H}^{1}(s)+\rho_{j,H}^{0}(s),
\end{eqnarray}
Then through dispersion relation and Borel transformation, we obtain the QCD sum rules at the hadron side,
\begin{eqnarray}
\notag\
\int_{m_{b}^{2}}^{s_{0}}\big[\sqrt{s}\rho_{j,H}^{1}(s)+\rho_{j,H}^{0}(s)\big]exp\big(-\frac{s}{T^2}\big)ds=2\lambda_{j}^{+2}M^{+}_{j}exp\big(-\frac{M^{+2}_{j}}{T^2}\big)+2\lambda_{j}^{\prime+2}M^{\prime+}_{j}exp\big(-\frac{M^{\prime+2}_{j}}{T^2}\big)
\end{eqnarray}
\begin{eqnarray}
\int_{m_{b}^{2}}^{s_{0}}\big[\sqrt{s}\rho_{j,H}^{1}(s)-\rho_{j,H}^{0}(s)\big]exp\big(-\frac{s}{T^2}\big)ds=2\lambda_{j}^{-2}M_{j}^{-}exp\big(-\frac{M^{-2}_{j}}{T^2}\big)+2\lambda_{j}^{\prime-2}M^{\prime-}_{j}exp\big(-\frac{M^{\prime-2}_{j}}{T^2}\big)
\end{eqnarray}
where $j$ denotes the total angular momentum $\frac{3}{2}$ or $\frac{5}{2}$, the subscript $H$ is for hadron side. The parameter $s_{0}$ is the continuum thresholds and $T^{2}$ are the Borel parameters. From Eq.(20), we can see that the bottom states with positive parity  and those with negative parity are successfully separated according to the combination of $\rho_{j,H}^{1}(s)$ and $\rho_{j,H}^{0}(s)$.

\begin{large}
\textbf{3.2 The QCD side}
\end{large}

At QCD side, the correlation function is approximated at
very large $P^2=-p^2$ by contract all quark fields with Wick's theorem. In our calculations, we use the full light quark propagators $S^{ij}_{q}(x)$ in the coordinate space, and the full heavy quark propagator $S^{ij}_{Q}(x)$ in the momentum spaces,
\begin{eqnarray}
 S_{q}^{ij}(x)=&&i\frac{x\!\!\!/}{2\pi^{2}x^{4}}\delta_{ij}-\frac{m_{q}}{4\pi^{2}x^{2}}\delta_{ij}-\frac{\langle
 \overline{q}q\rangle}{12}\Big(1-i\frac{m_{q}}{4}x\!\!\!/\Big)-\frac{x^{2}}{192}m_{0}^{2}\langle
 \overline{q}q\rangle\Big( 1-i\frac{m_{q}}{6}x\!\!\!/\Big)\\
 &&
 \notag-t^{a}_{ij}\Big[x\!\!\!/\sigma^{\theta\eta}+\sigma^{\theta\eta}x\!\!\!/\Big]\frac{i}{32\pi^{2}x^{2}}g_{s}G^{a}_{\theta\eta}
 -\frac{1}{8}\langle \overline{q}_{j}\sigma^{\mu\nu}q_{i}\rangle\sigma_{\mu\nu}\cdots,
\end{eqnarray}
\begin{eqnarray}
\notag
 S_{Q}^{ij}(x)=&&\frac{i}{(2\pi)^{4}}\int
 d^{4}ke^{-ik.x}\Big\{\frac{\delta_{ij}}{k\!\!\!/-m_{Q}}-\frac{g_{s}G_{\alpha\beta}^{c}t^{c}_{ij}}{4}\frac{\sigma^{\alpha\beta}(k\!\!\!/+m_{Q})+(k\!\!\!/+m_{Q})\sigma^{\alpha\beta}}{(k^{2}-m_{Q}^{2})^{2}}\\
 &&-\frac{g_{s}^{2}(t^{a}t^{b})_{ij}G_{\alpha\beta}^{a}G_{\mu\nu}^{b}(f^{\alpha\beta\mu\nu}+f^{\alpha\mu\beta\nu}+f^{\alpha\mu\nu\beta})}{4(k^{2}-m_{Q}^{2})^{5}}
 \cdots\Big\},
\end{eqnarray}
where
\begin{eqnarray}
f^{\alpha\beta\mu\nu}=(k\!\!\!/+m_{Q})\gamma^{\alpha}(k\!\!\!/+m_{Q})\gamma^{\beta}(k\!\!\!/+m_{Q})\gamma^{\mu}(k\!\!\!/+m_{Q})\gamma^{\nu}(k\!\!\!/+m_{Q}),
\end{eqnarray}
$q=u,d,s$, $t^{a}=\frac{\lambda^{a}}{2}$, the $\lambda^{a}$ is the Gell-Mann matrix.
After completing the integrals both in the coordinate and momentum spaces, we obtain the QCD spectral density through the imaginary part of the correlation,
\begin{eqnarray}
\frac{Im\Pi_{j}(s)}{\pi}= p\!\!\!/\rho_{j,QCD}^{1}(s)+\rho_{j,QCD}^{0}(s),
\end{eqnarray}
In calculations, we find the condensate contributions mainly come from the $\langle\overline{q}q\rangle$, $\langle\overline{s}s\rangle$, $\langle\frac{\alpha_{s}GG}{\pi}\rangle$, $\langle\overline{q}g_{s}\sigma Gq\rangle$, $\langle\overline{s}g_{s}\sigma Gs\rangle$, $\langle\overline{q}g_{s}\sigma Gq\rangle^{2}$, $\langle\overline{q}g_{s}\sigma Gq\rangle$$\langle\overline{s}g_{s}\sigma Gs\rangle$. The explicit form of the QCD spectral densities $\rho_{j,QCD}^{1}(s)$ and $\rho_{j,QCD}^{0}(s)$ are listed in the Appendix. It is the same as the hadron side, we can obtain the sum rules at the QCD side. Then, we take the quark-hadron duality below the continuum thresholds $s_{0}$ to obtain the QCD sum rules:
\begin{eqnarray}
\notag
2M_{j}^{+}\lambda_{j}^{+2}exp\big(-\frac{M^{+2}_{j}}{T^2}\big)+2M^{\prime+}_{j}\lambda_{j}^{\prime+2}exp\big(-\frac{M^{\prime+2}_{j}}{T^2}\big)=\int_{m_{b}^{2}}^{s_{0}}\big[\sqrt{s}\rho_{j,QCD}^{1}(s)+\rho_{j,QCD}^{0}(s)\big]exp\big(-\frac{s}{T^2}\big)ds \\
\end{eqnarray}
Firstly, we choose low continuum threshold parameters $s_{0}$ so as to include only the contributions of the $1D$ state. Then, we differentiate Eq.(25) with respect to $\frac{1}{T^{2}}$ to obtain the masses of the $1D$ $\Xi_{b}$ and $\Lambda_{b}$ states with $J^{p}=\frac{3}{2}^{+}$ and $\frac{5}{2}^{+}$,
\begin{eqnarray}
M_{j}^{+2}=\frac{-\frac{d}{d(1/T^{2})}\int_{m_{b}^{2}}^{s_{0}}\Big[\sqrt{s}\rho_{j,QCD}^{1}(s)+\rho_{j,QCD}^{0}(s)\Big]exp\big(-\frac{s}{T^2}\big)ds}{\int_{m_{b}^{2}}^{s_{0}}\Big[\sqrt{s}\rho_{j,QCD}^{1}(s)+\rho_{j,QCD}^{0}(s)\Big]exp\big(-\frac{s}{T^2}\big)ds}
\end{eqnarray}
After the mass $M_{j}^{+}$ is obtained, it is treated as a input parameter to obtain the pole residues,
\begin{eqnarray}
\lambda_{j}^{+2}=\frac{\int_{m_{b}^{2}}^{s_{0}}\Big[\sqrt{s}\rho_{j,QCD}^{1}(s)+\rho_{j,QCD}^{0}(s)\Big]exp\big(-\frac{s}{T^2}\big)ds}{2M_{+}exp\big(-\frac{M_{+}^{2}}{T^2}\big)ds}
\end{eqnarray}
Now, we take the masses and pole residues of the $1D$ states as input parameters, and postpone the continuum threshold parameters $s_{0}$ to larger values to include the contributions of the $2D$ states, and obtain the QCD sum rules for the masses and pole residues of the $2D$ states,
\begin{eqnarray}
M_{j}^{\prime+2}=\frac{-\frac{d}{d(1/T^{2})}\Big\{\int_{m_{b}^{2}}^{s^{\prime}_{0}}\Big[\sqrt{s}\rho_{j,QCD}^{1}(s)+\rho_{j,QCD}^{0}(s)\Big]exp\big(-\frac{s}{T^2}\big)ds-2M_{j}^{+}\lambda_{j}^{+2}exp\big(-\frac{M^{+2}_{j}}{T^2}\big)\Big\}\ }{\int_{m_{b}^{2}}^{s_{0}}\Big[\sqrt{s}\rho_{j,QCD}^{1}(s)+\rho_{j,QCD}^{0}(s)\Big]exp\big(-\frac{s}{T^2}\big)ds-2M_{j}^{+}\lambda_{j}^{+2}exp\big(-\frac{M^{+2}_{j}}{T^2}\big)}
\end{eqnarray}
\begin{eqnarray}
\lambda_{j}^{\prime+2}=\frac{\int_{m_{b}^{2}}^{s^{\prime}_{0}}\Big[\sqrt{s}\rho_{j,QCD}^{1}(s)+\rho_{j,QCD}^{0}(s)\Big]exp\big(-\frac{s}{T^2}\big)ds-2M_{j}^{+}\lambda_{j}^{+2}exp\big(-\frac{M^{+2}_{j}}{T^2}\big)}{2M^{\prime+}_{j}exp\big(-\frac{M_{j}^{\prime+2}}{T^2}\big)ds}
\end{eqnarray}
\begin{large}
\textbf{4 Numerical results and Discussions}
\end{large}

The calculated results from QCD sum rules depend on input parameters such as the vacuum condensates, the masses of quarks, the continuum threshold $s_{0}$ and Borel paramters $T^{2}$. For the values of the vacuum condensates using in this paper, we first take the standard values at the energy scale $\mu=1$ GeV\cite{parameters1,parameters2},
\begin{center}
$\langle \overline{q}q\rangle=-(0.24\pm0.01$ GeV)$^{3}$,

$\langle \overline{s}s\rangle=(0.8\pm0.1)\langle \overline{q}q\rangle$,

$\langle \overline{q}g_{s}\sigma Gq\rangle=m_{0}^{2}\langle \overline{q}q\rangle$,

$\langle \overline{s}g_{s}\sigma Gs\rangle=m_{0}^{2}\langle \overline{s}s\rangle$,

$m_{0}^{2}=(0.8\pm0.1)$GeV$^{2}$,

$\langle\frac{\alpha_{s}GG}{\pi}\rangle=$($0.33$GeV)$^{4}$
\end{center}
As for the masses of quarks, we set $m_{u}=m_{d}=0$ due to their small current quark masses and the masses of b-quark and s-quark are chosen to be
$m_{b}(m_{b})$=($4.18\pm0.03$)GeV and $m_{s}$($\mu=2$GeV)=($0.095\pm0.005$)GeV\cite{Xi5797}. Then, we take into account the energy-scale dependence of these above input parameters from the re-normalization group equation,
\begin{eqnarray}
\notag
\langle \overline{q}q\rangle(\mu)&&=\langle \overline{q}q\rangle(Q)\Big[\frac{\alpha_{s}(Q)}{\alpha_{s}(\mu)}\Big]^{\frac{4}{9}} \\
\notag
\langle \overline{s}s\rangle(\mu)&&=\langle \overline{s}s\rangle(Q)\Big[\frac{\alpha_{s}(Q)}{\alpha_{s}(\mu)}\Big]^{\frac{4}{9}} \\
\notag
\langle \overline{q}g_{s}\sigma Gq\rangle(\mu)&&=\langle \overline{q}g_{s}\sigma Gq\rangle(Q)\Big[\frac{\alpha_{s}(Q)}{\alpha_{s}(\mu)}\Big]^{\frac{2}{27}} \\
\notag
\langle \overline{s}g_{s}\sigma Gs\rangle(\mu)&&=\langle \overline{s}g_{s}\sigma Gs\rangle(Q)\Big[\frac{\alpha_{s}(Q)}{\alpha_{s}(\mu)}\Big]^{\frac{2}{27}} \\
\notag
m_{b}(\mu)&&=m_{b}(m_{b})\Big[\frac{\alpha_{s}(\mu)}{\alpha_{s}(m_{b})}\Big]^{\frac{12}{23}} \\
\notag
m_{s}(\mu)&&=m_{s}(2GeV)\Big[\frac{\alpha_{s}(\mu)}{\alpha_{s}(2GeV)}\Big]^{\frac{4}{9}} \\
\notag
\alpha_{s}(\mu)&&=\frac{1}{b_{0}t}\Big[ 1-\frac{b_{1}}{b_{0}^{2}}\frac{logt}{t}+\frac{b_{1}^{2}(log^{2}t-logt-1)+b_{0}b_{2}}{b_{0}^{4}t^{2}}\Big]
\end{eqnarray}
where $t$=log$\frac{\mu^{2}}{\Lambda^{2}}$, $b_{0}=\frac{33-2n_{f}}{12\pi}$, $b_{1}=\frac{153-19n_{f}}{24\pi^{2}}$, $b_{2}=\frac{2857-\frac{5033}{9}n_{f}+\frac{325}{27}n_{f}^{2}}{128\pi^{3}}$, $\Lambda=213$ MeV, $296$ MeV, $339$ MeV for the flavors $n_{f}=5$, $4$ and $3$, respectively\cite{Xi5797}, and evolve these parameters to the optimal energy scales $\mu$ to extract the masses of the bottom baryon states. In order to determine the optimal energy scales, we have developed an empirical formula $\mu=\sqrt{M_{H}^{2}-(n\mathbb{M}_{Q})^{2}}$, where $M_{H}$ is the mass of hadron, $\mathbb{M}_{Q}$ is the effective mass of heavy quark, and $n$ stands for the number of heavy quarks within a hadron. Since this formula was proposed to determine the optimal energy scales $\mu$ in the calculations of QCD sum rules\cite{energy1,energy2, energy3}, it has successfully been used to study the hidden-charm(hidden-bottom) tetraquark states and molecular states\cite{energy1,energy2, energy3}, hidden-charm pentaquark states\cite{energy4}, charmed and bottom states\cite{energy5}, etc. In this article, we choose the effective mass of b-quark to be $\mathbb{M}_{b}=5.17$ GeV which was fitted in the study of the diquark-antidiquark type hidden-bottom tetraquark states\cite{energy6}.

In order to choose the working interval of the parameters $T^{2}$ and
continuum threshold parameters $s_{0}$, some criteria should be satisfied, which are pole dominance, convergence of operator production
expansion(OPE), appearance of the Borel platforms and satisfying the energy scale formula. That is to say, the pole contribution should be as large as possible(commonly larger than $40\%$) comparing with the contributions of the high resonances and continuum states. Meanwhile, we should also find a plateau(Borel platforms), which will ensure OPE convergence and the stability
of the final results. The plateau is often called Borel window.

After repeated adjustment and comparison, we finally determine the the optimal energy scales $\mu$, the Borel windows,
the continuum threshold parameters $s_{0}$ and the pole contributions, which are presented in Tables I-II.  As an example, the results for $1D$ states with different excitation modes are shown explicitly in Figs1-24. It should be noticed that we plot the masses and pole residues with variations of the Borel parameters at much larger intervals than the Borel windows shown in Tables I-II. And the uncertainties of the masses and pole residues are marked as the Upper bound and Lower bound in these figures. From Tables I-II, we observe that the pole contributions are about ($40-80$)$\%$, the pole dominance criterion is satisfied. On the other hand, we can see that there appear flat platforms in Figs.1-24, the uncertainties originating from the Borel parameters $T^2$ in the Borel window are small($\leq3 \%$). That is to say, all of the criteria of QCD sum rules are satisfied, it is reliable to extract the final results about the D-wave bottom baryons. Taking into account all uncertainties of the input parameters, we obtain the masses and pole residues of $1D$ and $2D$ states of $\Lambda_{b}$ and $\Xi_{b}$ baryons, which are also presentd in Tables I-II.
\begin{table*}[htbp]
\begin{ruledtabular}\caption{The optimal energy scales $\mu$, Borel parameters $T^{2}$
, continuum threshold parameters $s_{0}$, pole contributions (pole) and the masses, pole residues for the
D-wave bottom baryon states $\Xi_{b}$, where the results of Ref.[15] are the quark-model predictions.}
\begin{tabular}{c| c c c c c c c c c c}
 $\Xi_{b}$$(L_{\rho},l_{\lambda})$ &\ $J^{P}$ & \ $\mu$(GeV$^{2}$) &\ $T^{2}$(GeV$^{2}$) &\ $\sqrt{s_{0}}$(GeV)  &\ M(GeV) &\ Exp(GeV) &\ $\lambda$($10^{-1}$GeV$^{5}$) &\ pole     \\
 \hline
$\Xi_{b}$($2,0$)  &\  $\frac{5}{2}^{+}$($1$D)  &\ $3.7$ &\  $3.8-4.2$   &\ $7.0\pm0.1$ &\ $6.43^{+0.10}_{-0.10}$ &\   &\ $1.59_{-0.18}^{+0.20}$ &\ $(49-59)\%$ \\
$\Xi_{b}$($2,0$)  &\  $\frac{3}{2}^{+}$($1$D)   &\ $3.5$ &\  $3.9-4.3$   &\ $7.0\pm0.1$ &\ $6.42^{+0.09}_{-0.09}$ &\  &\ $4.64^{+0.60}_{-0.59}$ &\ $(47-57)\%$ \\
$\Xi_{b}$($0,2$)  &\  $\frac{5}{2}^{+}$($1$D)   &\ $3.6$ &\  $4.3-4.7$   &\ $6.9\pm0.1$ &\ $6.36^{+0.11}_{-0.12}$ &\ $6.333$\cite{Xi6333} &\ $0.67^{+0.08}_{-0.07}$ &\ $(41-56)\%$ \\
$\Xi_{b}$($0,2$)  &\  $\frac{3}{2}^{+}$($1$D)   &\ $3.6$ &\  $3.6-4.0$   &\ $6.9\pm0.1$ &\ $6.34^{+0.12}_{-0.11}$ &\  $6.327$\cite{Xi6333} &\ $2.98^{+0.38}_{-0.32}$ &\ $(41-61)\%$ \\
$\Xi_{b}$($1,1$)  &\  $\frac{5}{2}^{+}$($1$D)   &\ $3.6$ &\  $4.2-4.6$   &\ $6.9\pm0.1$ &\ $6.41^{+0.09}_{-0.11}$ &\  &\ $0.80^{+0.11}_{-0.12}$ &\ $(42-58)\%$ \\
$\Xi_{b}$($1,1$)  &\  $\frac{3}{2}^{+}$($1$D)   &\ $3.7$ &\  $3.8-4.2$   &\ $7.0\pm0.1$ &\ $6.41^{+0.09}_{-0.11}$ &\  &\ $2.82^{+0.30}_{-0.32}$ &\ $(47-57)\%$ \\
$\Xi_{b}$($2,0$)  &\  $\frac{5}{2}^{+}$($2$D)   &\ $4.1$ &\  $3.9-4.3$   &\ $7.3\pm0.1$ &\ $6.77^{+0.12}_{-0.11}$ &\   &\ $2.46_{-0.19}^{+0.23}$ &\ $(65-77)\%$ \\
$\Xi_{b}$($2,0$)  &\  $\frac{3}{2}^{+}$($2$D)   &\ $4.2$ &\  $3.9-4.3$   &\ $7.3\pm0.1$ &\ $6.73^{+0.09}_{-0.10}$ &\  &\ $7.13^{+0.55}_{-0.60}$ &\ $(66-76)\%$ \\
$\Xi_{b}$($0,2$)  &\  $\frac{5}{2}^{+}$($2$D)   &\ $4.1$ &\  $4.3-4.7$   &\ $7.2\pm0.1$ &\ $6.69^{+0.13}_{-0.11}$ &\ $6.696$\cite{quam6}  &\ $0.98^{+0.10}_{-0.12}$ &\ $(60-68)\%$ \\
$\Xi_{b}$($0,2$)  &\  $\frac{3}{2}^{+}$($2$D)   &\ $4.1$ &\  $3.7-4.1$   &\ $7.2\pm0.1$ &\ $6.62^{+0.10}_{-0.13}$ &\ $6.690$\cite{quam6}  &\ $4.29^{+0.42}_{-0.38}$ &\ $(53-75)\%$ \\
$\Xi_{b}$($1,1$)  &\  $\frac{5}{2}^{+}$($2$D)   &\ $4.1$ &\  $4.2-4.6$   &\ $7.2\pm0.1$ &\ $6.72^{+0.11}_{-0.13}$ &\  &\ $1.19^{+0.13}_{-0.15}$ &\ $(57-70)\%$ \\
$\Xi_{b}$($1,1$)  &\  $\frac{3}{2}^{+}$($2$D)   &\ $4.1$ &\  $3.8-4.2$   &\ $7.3\pm0.1$ &\ $6.79^{+0.12}_{-0.09}$ &\  &\ $3.53^{+0.35}_{-0.40}$ &\ $(65-78)\%$ \\
\end{tabular}
\end{ruledtabular}
\end{table*}
\begin{table*}[htbp]
\begin{ruledtabular}\caption{The optimal energy scales $\mu$, Borel parameters $T^{2}$
, continuum threshold parameters $s_{0}$, pole contributions (pole) and the masses, pole residues for the
D-wave bottom baryon states $\Lambda_{b}$, where the results of Ref.[15] are the quark-model predictions.}
\begin{tabular}{c| c c c c c c c c c}
 $\Lambda_{b}$$(L_{\rho},l_{\lambda})$ &\ $J^{P}$ & \ $\mu$(GeV$^{2}$) &\ $T^{2}$(GeV$^{2}$) &\ $\sqrt{s_{0}}$(GeV)  &\ M(GeV) &\ Exp(GeV) &\ $\lambda$($10^{-1}$GeV$^{5}$) &\ pole     \\
 \hline
$\Lambda_{b}$($2,0$)  &\  $\frac{5}{2}^{+}$($1$D)   &\ $3.2$ &\  $3.5-3.9$   &\ $6.8\pm0.1$ &\ $6.28^{+0.10}_{-0.10}$ &\  &\ $0.96^{+0.10}_{-0.13}$ &\ $(42-58)\%$ \\
$\Lambda_{b}$($2,0$)  &\  $\frac{3}{2}^{+}$($1$D)   &\ $3.2$ &\  $3.3-3.7$   &\ $6.7\pm0.1$ &\ $6.21^{+0.10}_{-0.10}$ &\  &\ $2.23^{+0.35}_{-0.33}$ &\ $(44-56)\%$ \\
$\Lambda_{b}$($0,2$)  &\  $\frac{5}{2}^{+}$($1$D)   &\ $3.2$ &\  $3.7-4.1$   &\ $6.7\pm0.1$ &\ $6.15^{+0.13}_{-0.15}$ &\ $6.153$\cite{LHCb2}  &\ $0.37^{+0.05}_{-0.04}$ &\ $(41-56)\%$ \\
$\Lambda_{b}$($0,2$)  &\  $\frac{3}{2}^{+}$($1$D)   &\ $3.2$ &\  $3.4-3.8$   &\ $6.6\pm0.1$ &\ $6.13^{+0.10}_{-0.09}$ &\ $6.146$\cite{LHCb2}  &\ $1.45^{+0.21}_{-0.22}$ &\ $(44-59)\%$ \\
$\Lambda_{b}$($1,1$)  &\  $\frac{5}{2}^{+}$($1$D)   &\ $3.2$ &\  $3.9-4.3$   &\ $6.8\pm0.1$ &\ $6.29^{+0.08}_{-0.06}$ &\  &\ $0.54^{+0.08}_{-0.09}$ &\ $(45-60)\%$ \\
$\Lambda_{b}$($1,1$)  &\  $\frac{3}{2}^{+}$($1$D)   &\ $3.4$ &\  $3.5-3.9$   &\ $6.8\pm0.1$ &\ $6.30^{+0.08}_{-0.07}$ &\  &\ $1.77^{+0.31}_{-0.28}$ &\ $(42-57)\%$ \\
$\Lambda_{b}$($2,0$)  &\  $\frac{5}{2}^{+}$($2$D)   &\ $3.9$ &\  $3.7-4.1$   &\ $7.1\pm0.1$ &\ $6.57^{+0.12}_{-0.11}$ &\  &\ $1.84^{+0.18}_{-0.20}$ &\ $(59-73)\%$ \\
$\Lambda_{b}$($2,0$)  &\  $\frac{3}{2}^{+}$($2$D)   &\ $3.9$ &\  $3.6-4.0$   &\ $7.0\pm0.1$ &\ $6.50^{+0.11}_{-0.11}$ &\  &\ $4.65^{+0.42}_{-0.38}$ &\ $(56-67)\%$ \\
$\Lambda_{b}$($0,2$)  &\  $\frac{5}{2}^{+}$($2$D)   &\ $3.8$ &\  $3.9-4.3$   &\ $7.0\pm0.1$ &\ $6.53^{+0.14}_{-0.14}$ &\  $6.531$\cite{quam6}  &\ $0.83^{+0.08}_{-0.07}$ &\ $(61-71)\%$ \\
$\Lambda_{b}$($0,2$)  &\  $\frac{3}{2}^{+}$($2$D)   &\ $3.8$ &\  $3.6-4.0$   &\ $6.9\pm0.1$ &\ $6.47^{+0.09}_{-0.10}$ &\ $6.526$\cite{quam6}  &\ $3.00^{+0.31}_{-0.30}$ &\ $(50-65)\%$ \\
$\Lambda_{b}$($1,1$)  &\  $\frac{5}{2}^{+}$($2$D)   &\ $3.9$ &\  $4.1-4.5$   &\ $7.1\pm0.1$ &\ $6.62^{+0.10}_{-0.08}$ &\  &\ $1.05^{+0.12}_{-0.13}$ &\ $(53-66)\%$ \\
$\Lambda_{b}$($1,1$)  &\  $\frac{3}{2}^{+}$($2$D)   &\ $3.9$ &\  $3.7-4.1$   &\ $7.1\pm0.1$ &\ $6.60^{+0.09}_{-0.09}$ &\  &\ $2.96^{+0.42}_{-0.39}$ &\ $(56-72)\%$ \\
\end{tabular}
\end{ruledtabular}
\end{table*}
\begin{figure}[h]
\begin{minipage}[t]{0.45\linewidth}
\centering
\includegraphics[height=5cm,width=7cm]{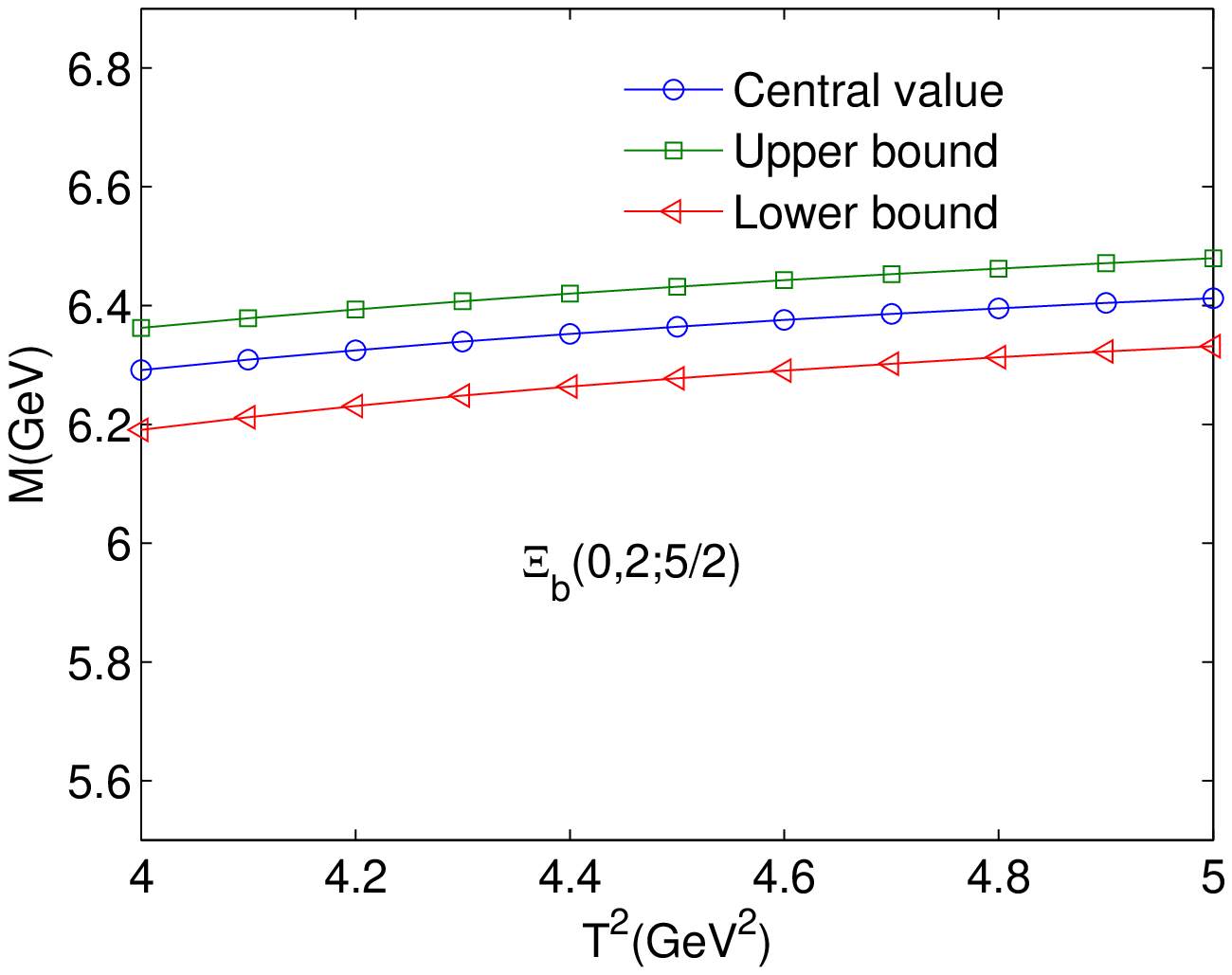}
\caption{The mass of the bottom baryon state $\Xi_{b}$(0,2,$\frac{5}{2}$) with variations of the Borel parameters $T^{2}$}
\end{minipage}
\hfill
\begin{minipage}[t]{0.45\linewidth}
\centering
\includegraphics[height=5cm,width=7cm]{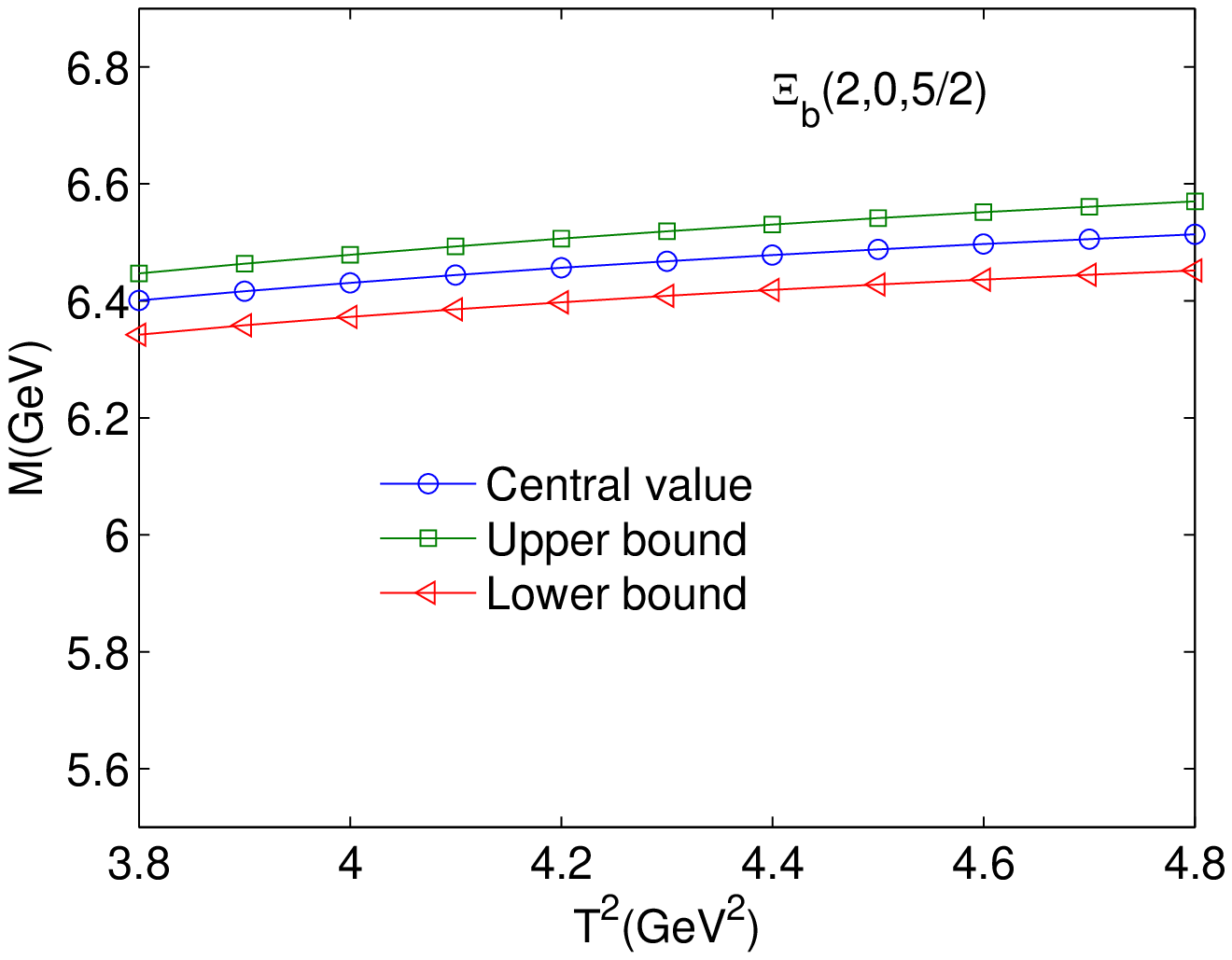}
\caption{The mass of the bottom baryon state $\Xi_{b}$(2,0,$\frac{5}{2}$) with variations of the Borel parameters $T^{2}$}
\end{minipage}
\end{figure}
\begin{figure}[h]
\begin{minipage}[t]{0.45\linewidth}
\centering
\includegraphics[height=5cm,width=7cm]{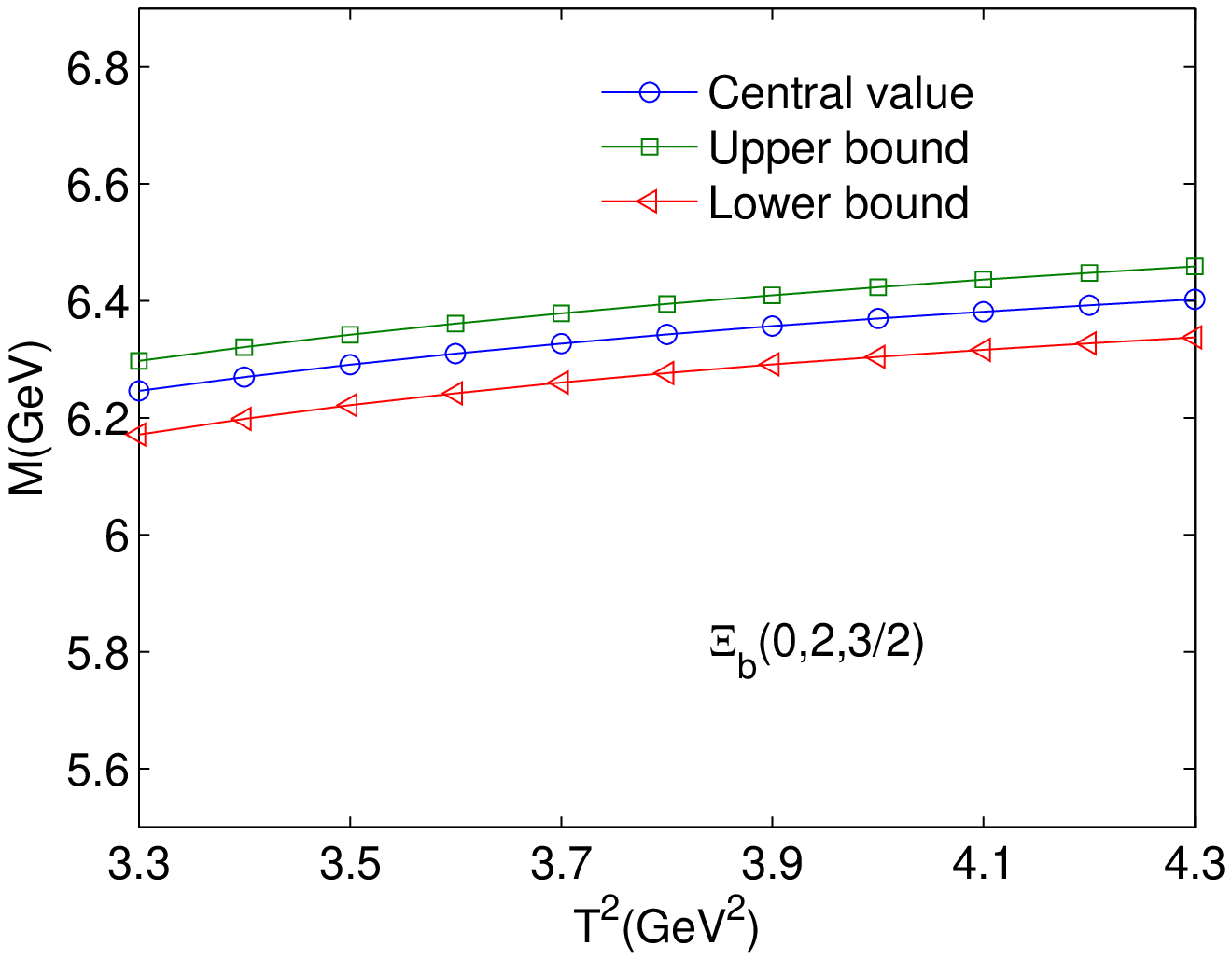}
\caption{The mass of the bottom baryon state $\Xi_{b}$(0,2,$\frac{3}{2}$) with variations of the Borel parameters $T^{2}$}
\end{minipage}
\hfill
\begin{minipage}[t]{0.45\linewidth}
\centering
\includegraphics[height=5cm,width=7cm]{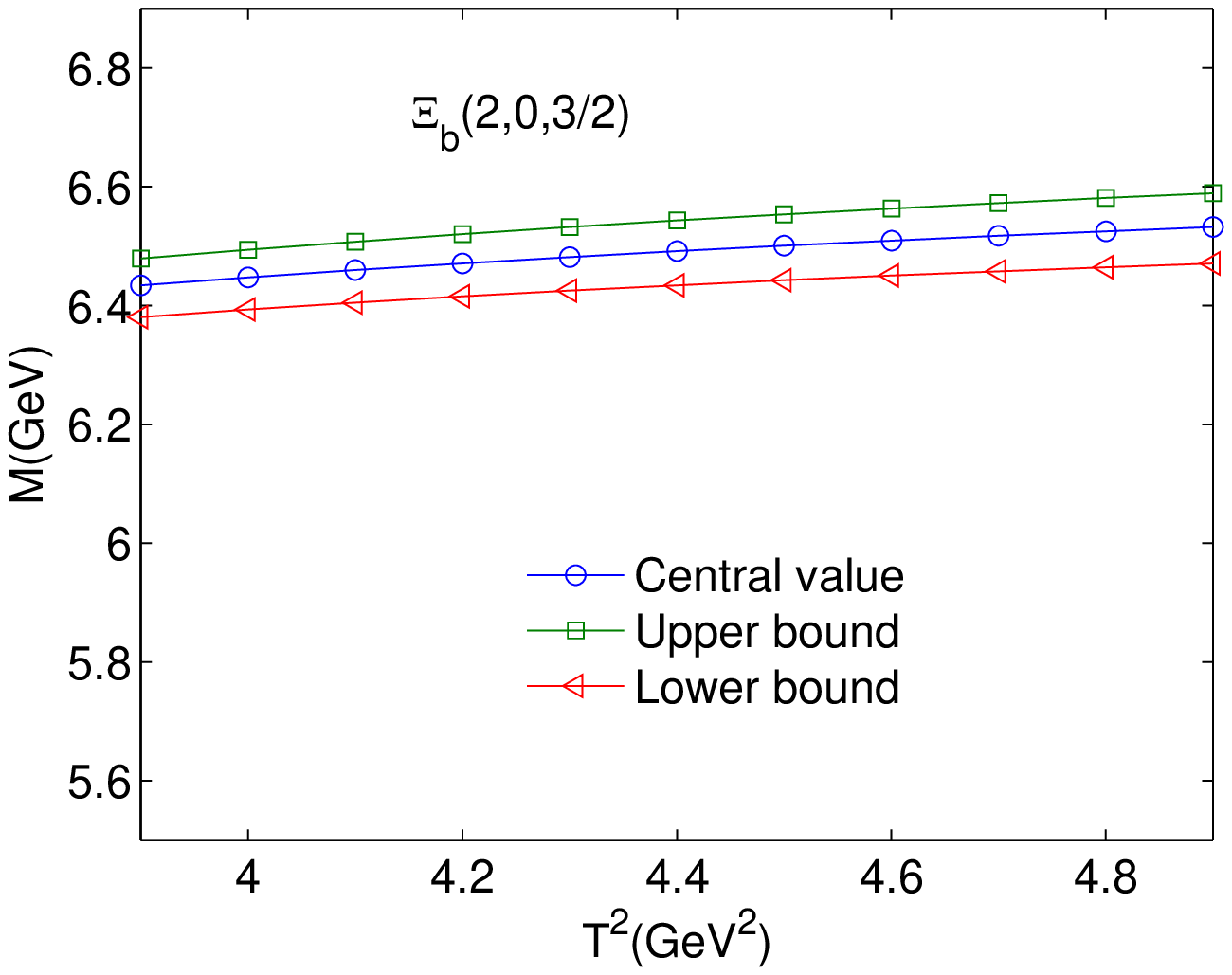}
\caption{The mass of the bottom baryon state $\Xi_{b}$(2,0,$\frac{3}{2}$) with variations of the Borel parameters $T^{2}$}
\end{minipage}
\end{figure}
\begin{figure}[h]
\begin{minipage}[t]{0.45\linewidth}
\centering
\includegraphics[height=5cm,width=7cm]{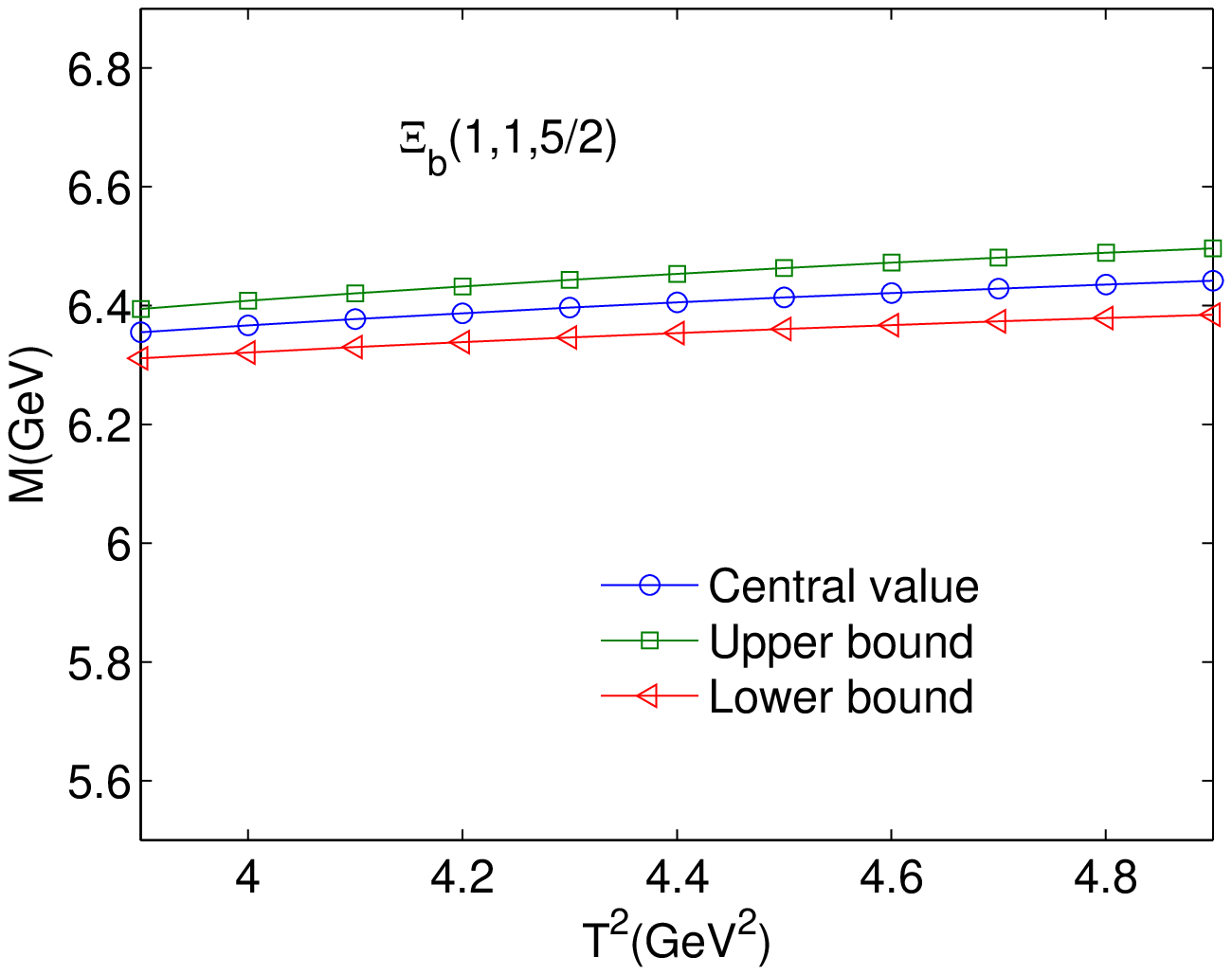}
\caption{The mass of the bottom baryon state $\Xi_{b}$(1,1,$\frac{5}{2}$) with variations of the Borel parameters $T^{2}$}
\end{minipage}
\hfill
\begin{minipage}[t]{0.45\linewidth}
\centering
\includegraphics[height=5cm,width=7cm]{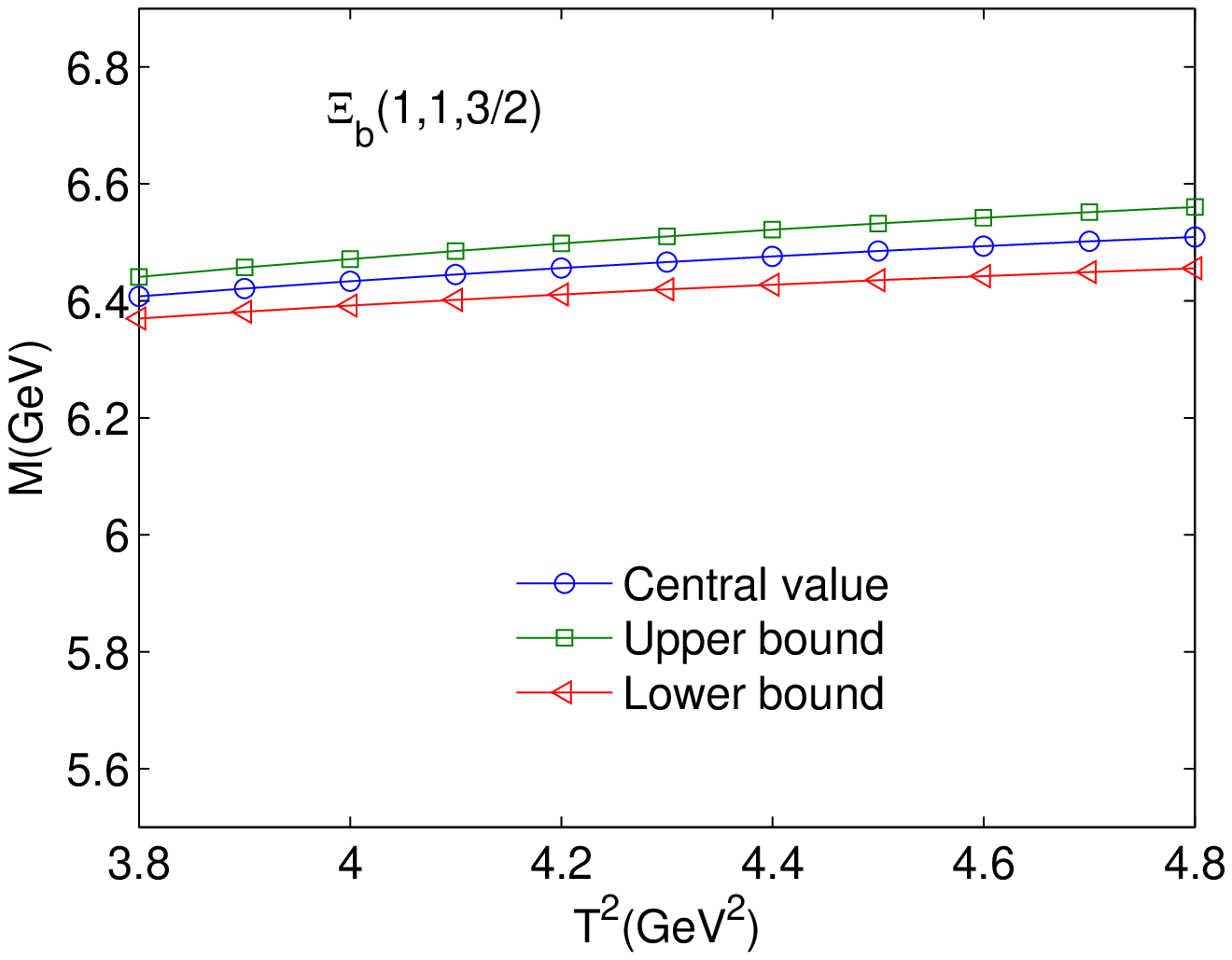}
\caption{The mass of the bottom baryon state $\Xi_{b}$(1,1,$\frac{3}{2}$) with variations of the Borel parameters $T^{2}$}
\end{minipage}
\end{figure}
\begin{figure}[h]
\begin{minipage}[t]{0.45\linewidth}
\centering
\includegraphics[height=5cm,width=7cm]{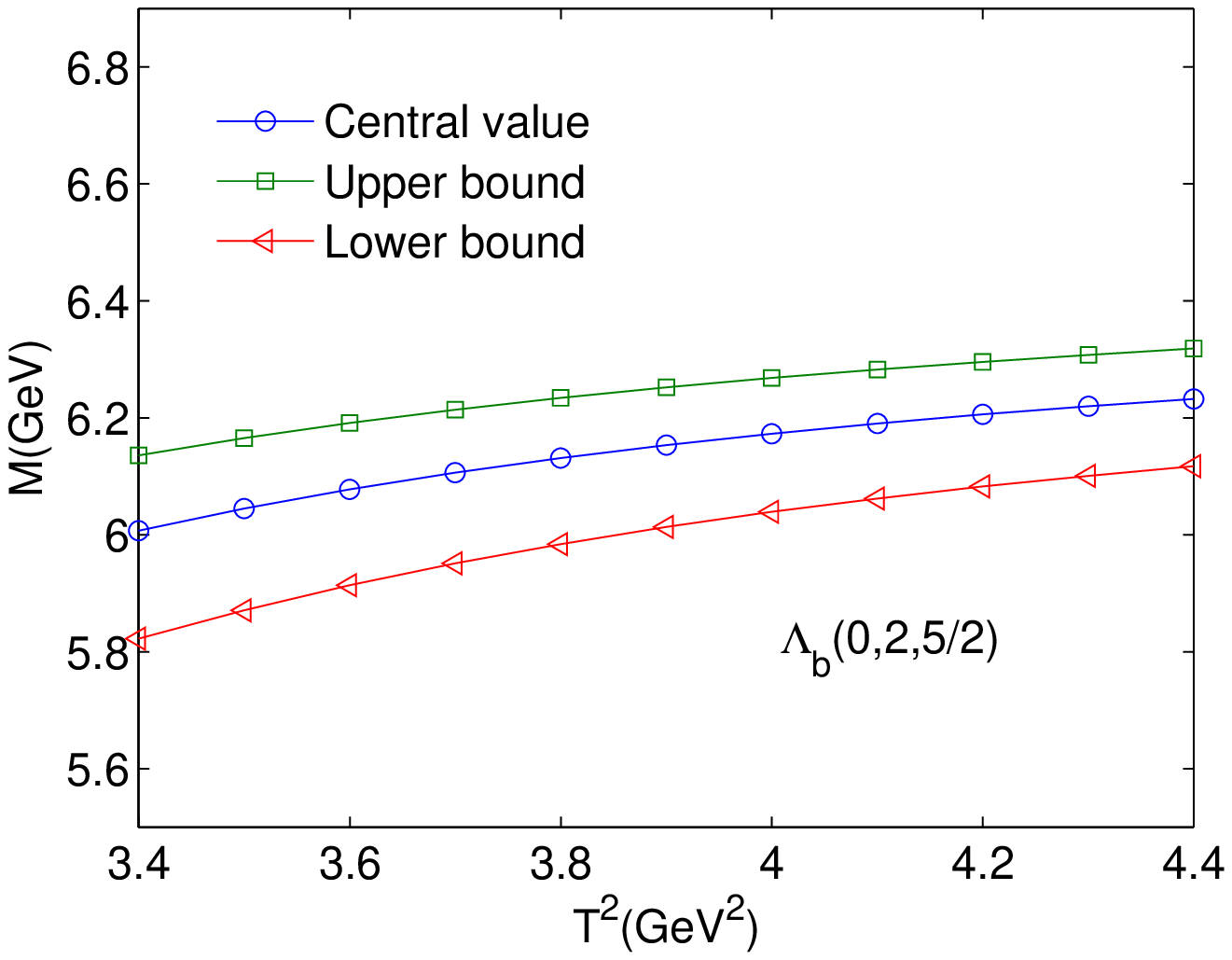}
\caption{The mass of the bottom baryon state $\Lambda_{b}$(0,2,$\frac{5}{2}$) with variations of the Borel parameters $T^{2}$}
\end{minipage}
\hfill
\begin{minipage}[t]{0.45\linewidth}
\centering
\includegraphics[height=5cm,width=7cm]{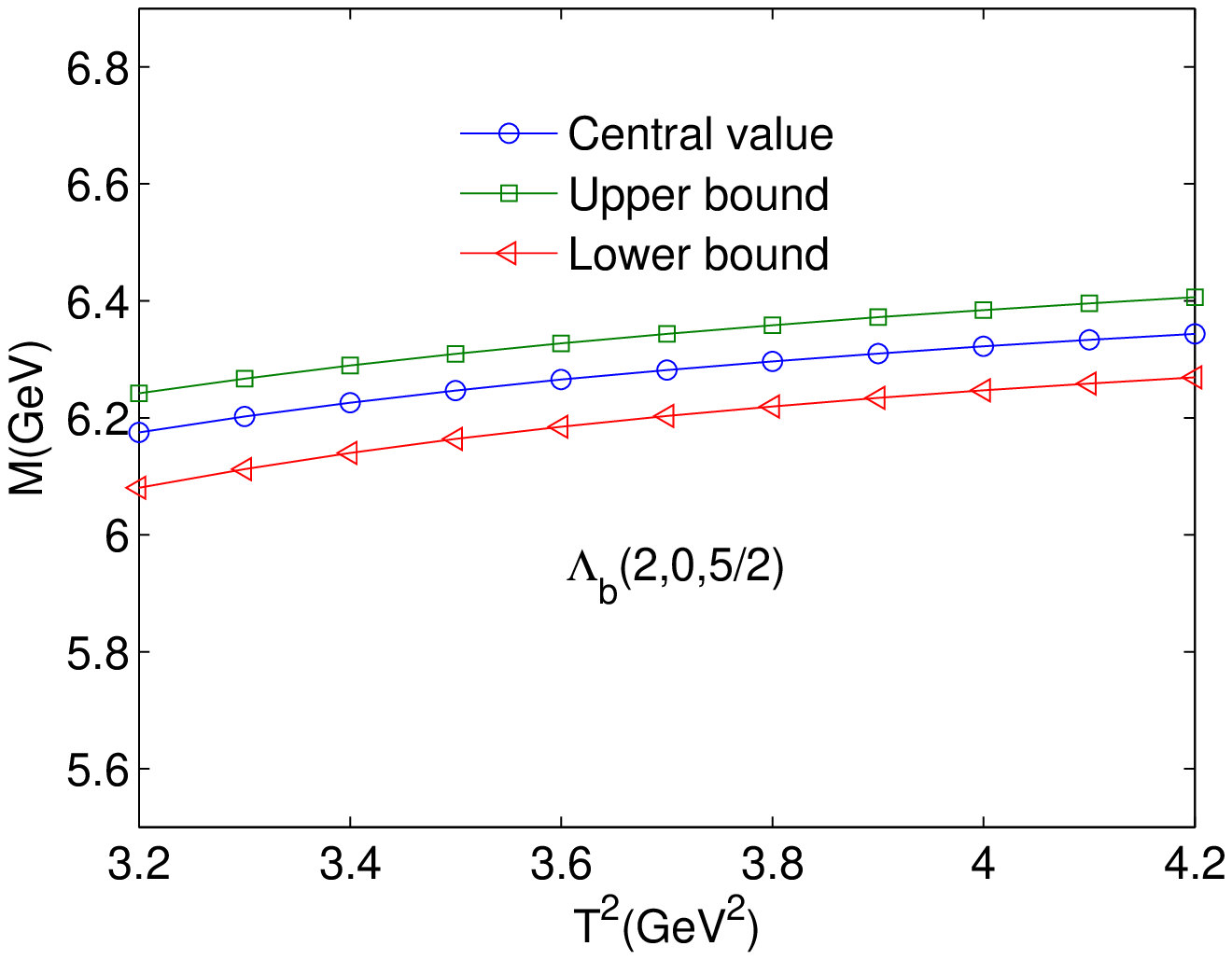}
\caption{The mass of the bottom baryon state $\Lambda_{b}$(2,0,$\frac{5}{2}$) with variations of the Borel parameters $T^{2}$}
\end{minipage}
\end{figure}
\begin{figure}[h]
\begin{minipage}[t]{0.45\linewidth}
\centering
\includegraphics[height=5cm,width=7cm]{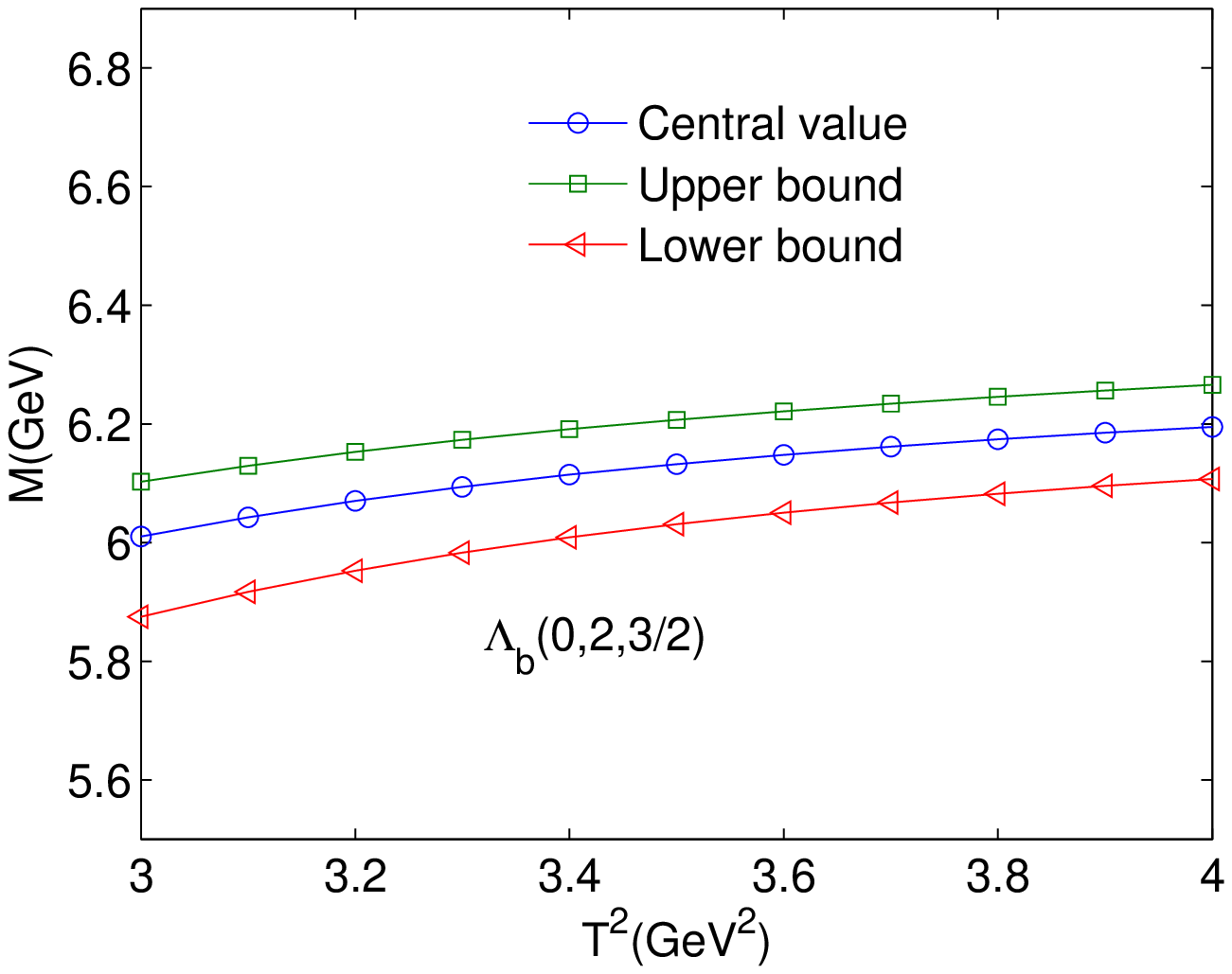}
\caption{The mass of the bottom baryon state $\Lambda_{b}$(0,2,$\frac{3}{2}$) with variations of the Borel parameters $T^{2}$}
\end{minipage}
\hfill
\begin{minipage}[t]{0.45\linewidth}
\centering
\includegraphics[height=5cm,width=7cm]{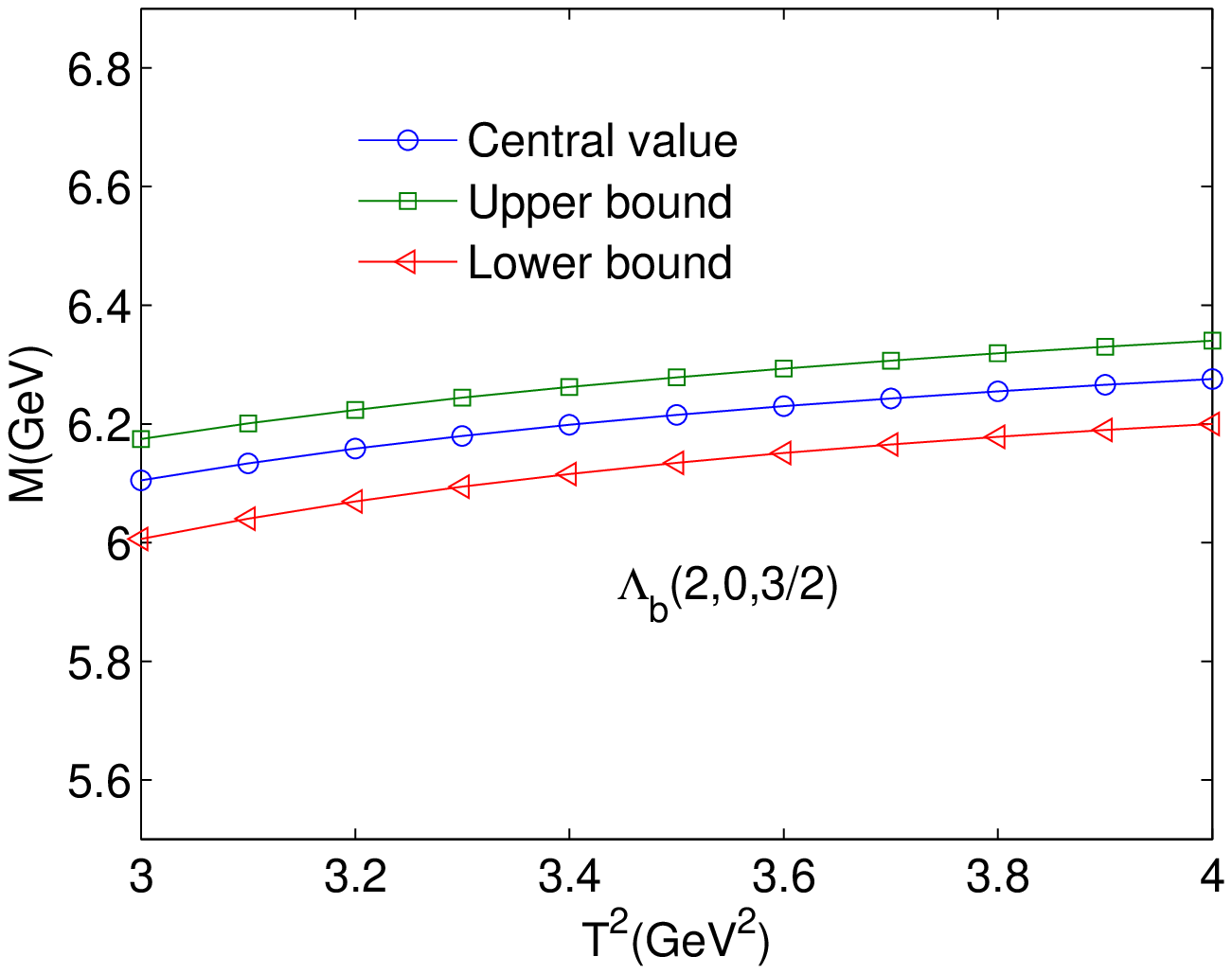}
\caption{The mass of the bottom baryon state $\Lambda_{b}$(2,0,$\frac{3}{2}$) with variations of the Borel parameters $T^{2}$}
\end{minipage}
\end{figure}
\begin{figure}[h]
\begin{minipage}[t]{0.45\linewidth}
\centering
\includegraphics[height=5cm,width=7cm]{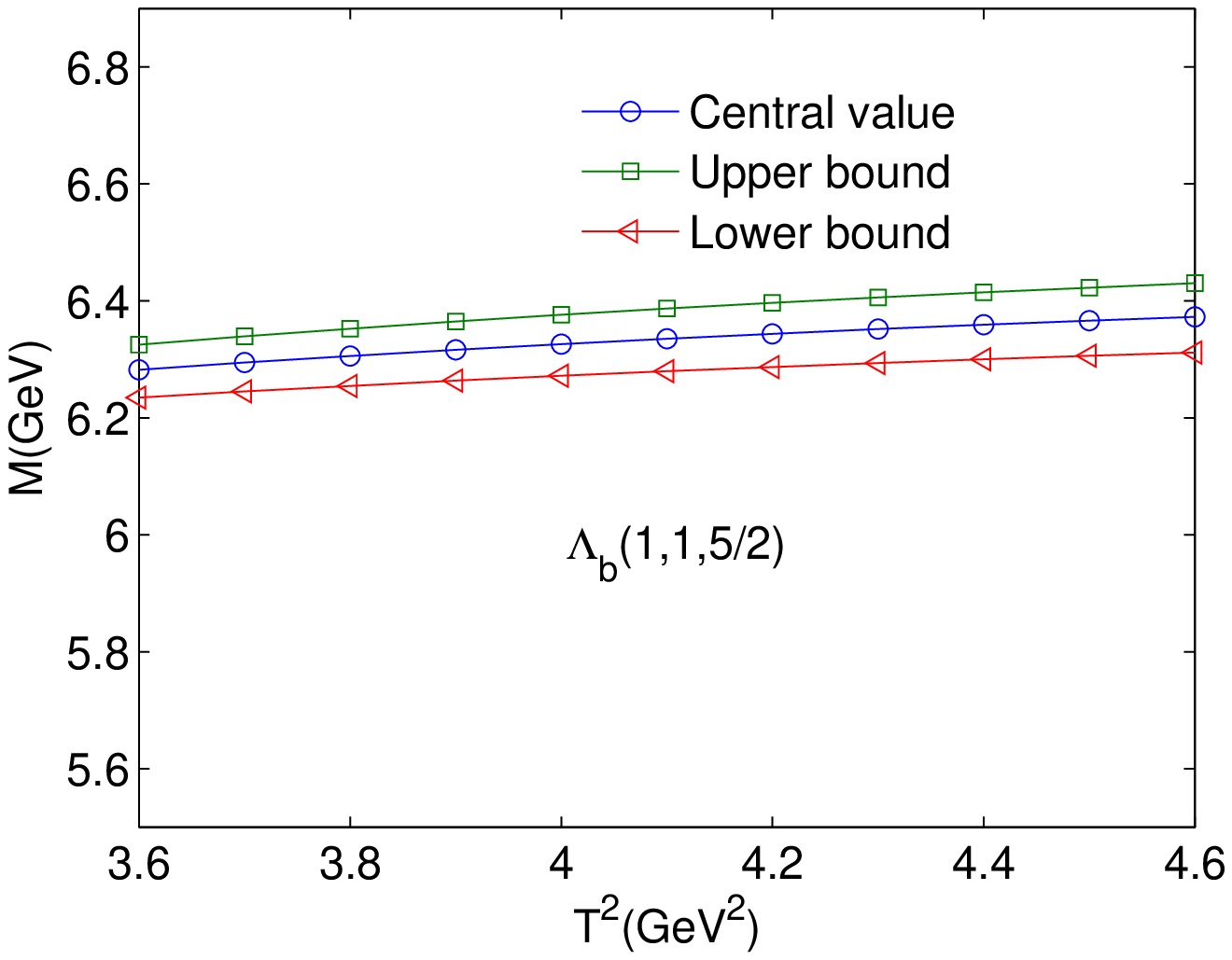}
\caption{The mass of the bottom baryon state $\Lambda_{b}$(1,1,$\frac{5}{2}$) with variations of the Borel parameters $T^{2}$}
\end{minipage}
\hfill
\begin{minipage}[t]{0.45\linewidth}
\centering
\includegraphics[height=5cm,width=7cm]{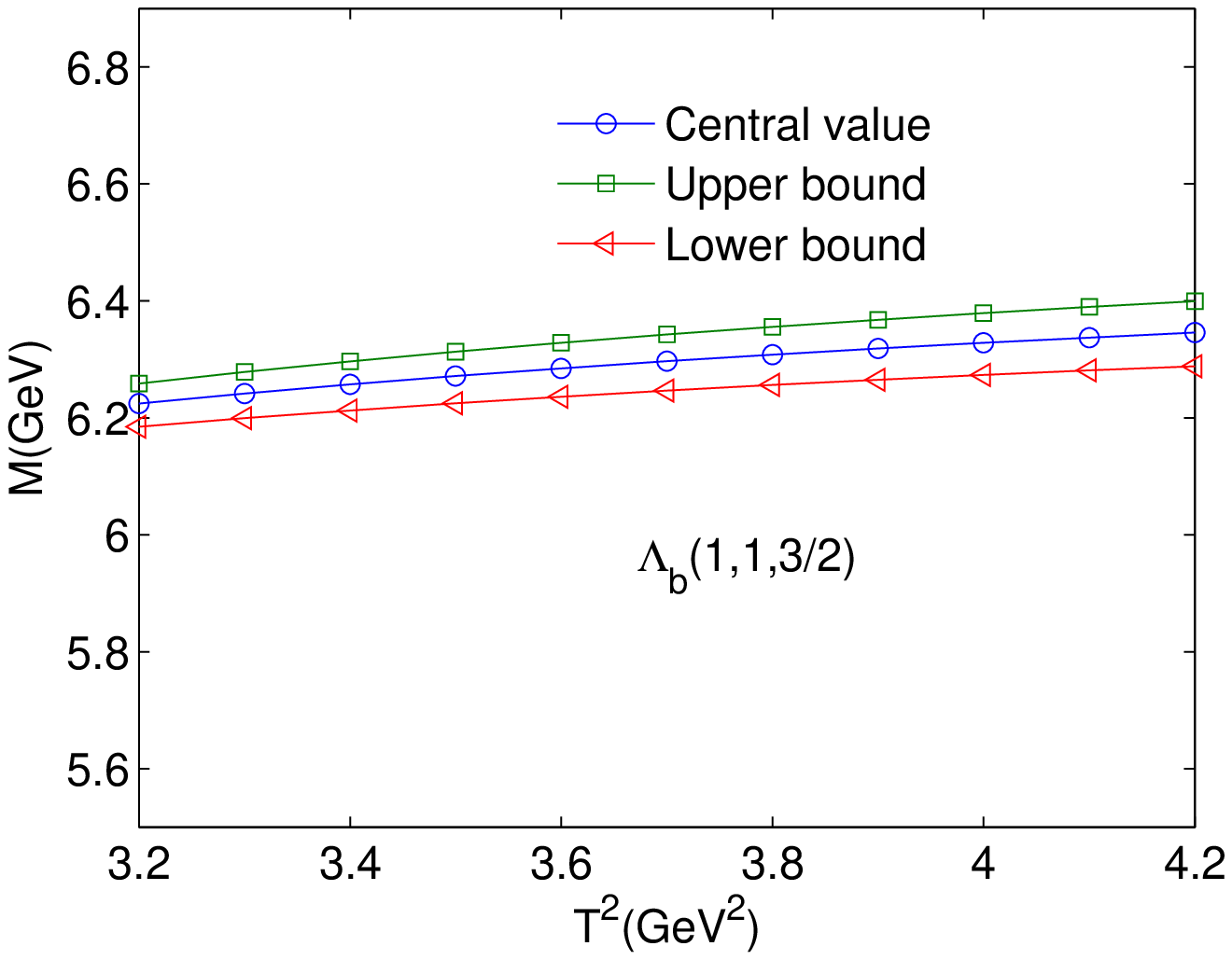}
\caption{The mass of the bottom baryon state $\Lambda_{b}$(1,1,$\frac{3}{2}$) with variations of the Borel parameters $T^{2}$}
\end{minipage}
\end{figure}
\begin{figure}[h]
\begin{minipage}[t]{0.45\linewidth}
\centering
\includegraphics[height=5cm,width=7cm]{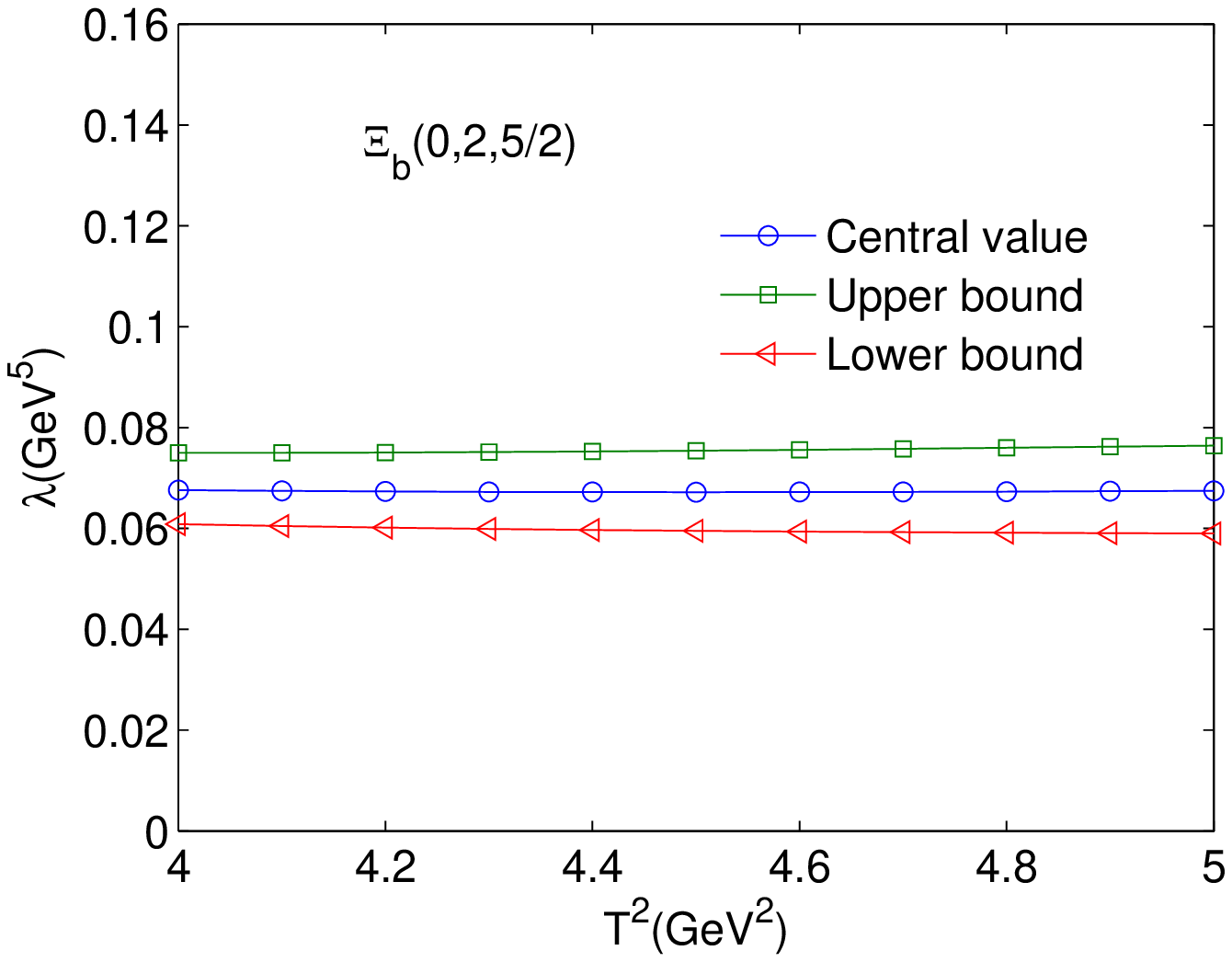}
\caption{The pole residues of the bottom baryon state $\Xi_{b}$(0,2,$\frac{5}{2}$) with variations of the Borel
parameters $T^{2}$}
\end{minipage}
\hfill
\begin{minipage}[t]{0.45\linewidth}
\centering
\includegraphics[height=5cm,width=7cm]{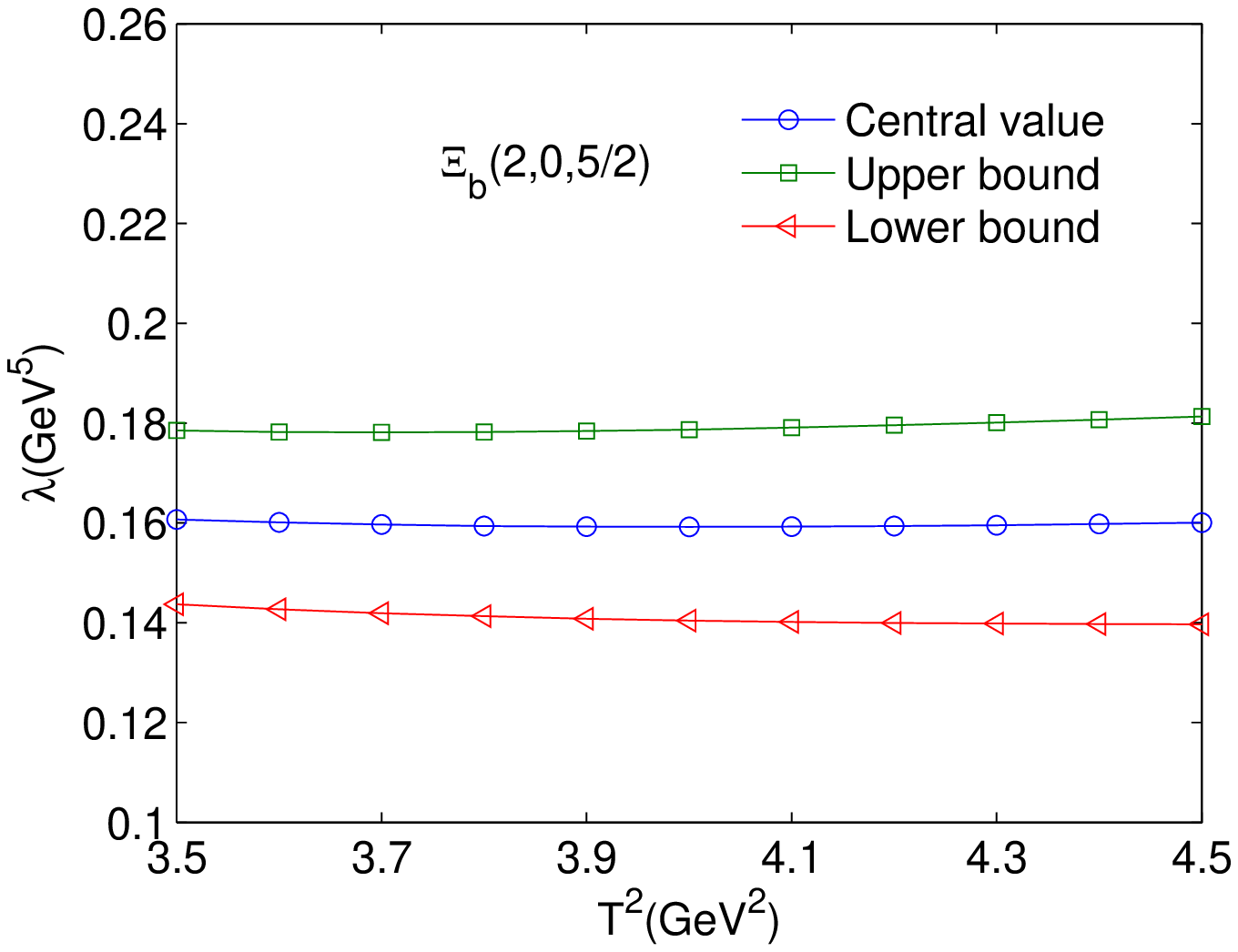}
\caption{The pole residues of the bottom baryon state $\Xi_{b}$(2,0,$\frac{5}{2}$) with variations of the Borel
parameters $T^{2}$}
\end{minipage}
\end{figure}
\begin{figure}[h]
\begin{minipage}[t]{0.45\linewidth}
\centering
\includegraphics[height=5cm,width=7cm]{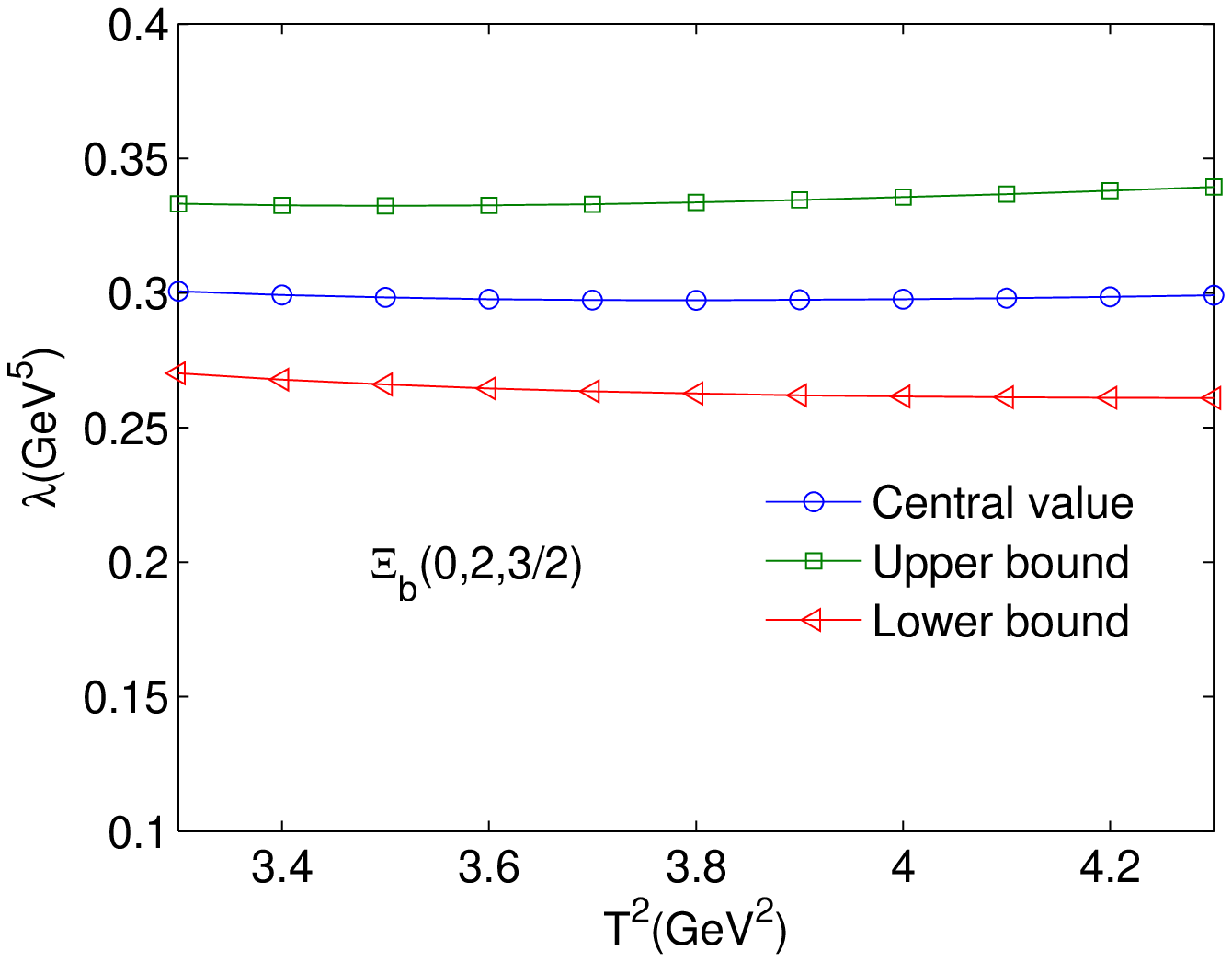}
\caption{The pole residues of the bottom baryon state $\Xi_{b}$(0,2,$\frac{3}{2}$) with variations of the Borel
parameters $T^{2}$}
\end{minipage}
\hfill
\begin{minipage}[t]{0.45\linewidth}
\centering
\includegraphics[height=5cm,width=7cm]{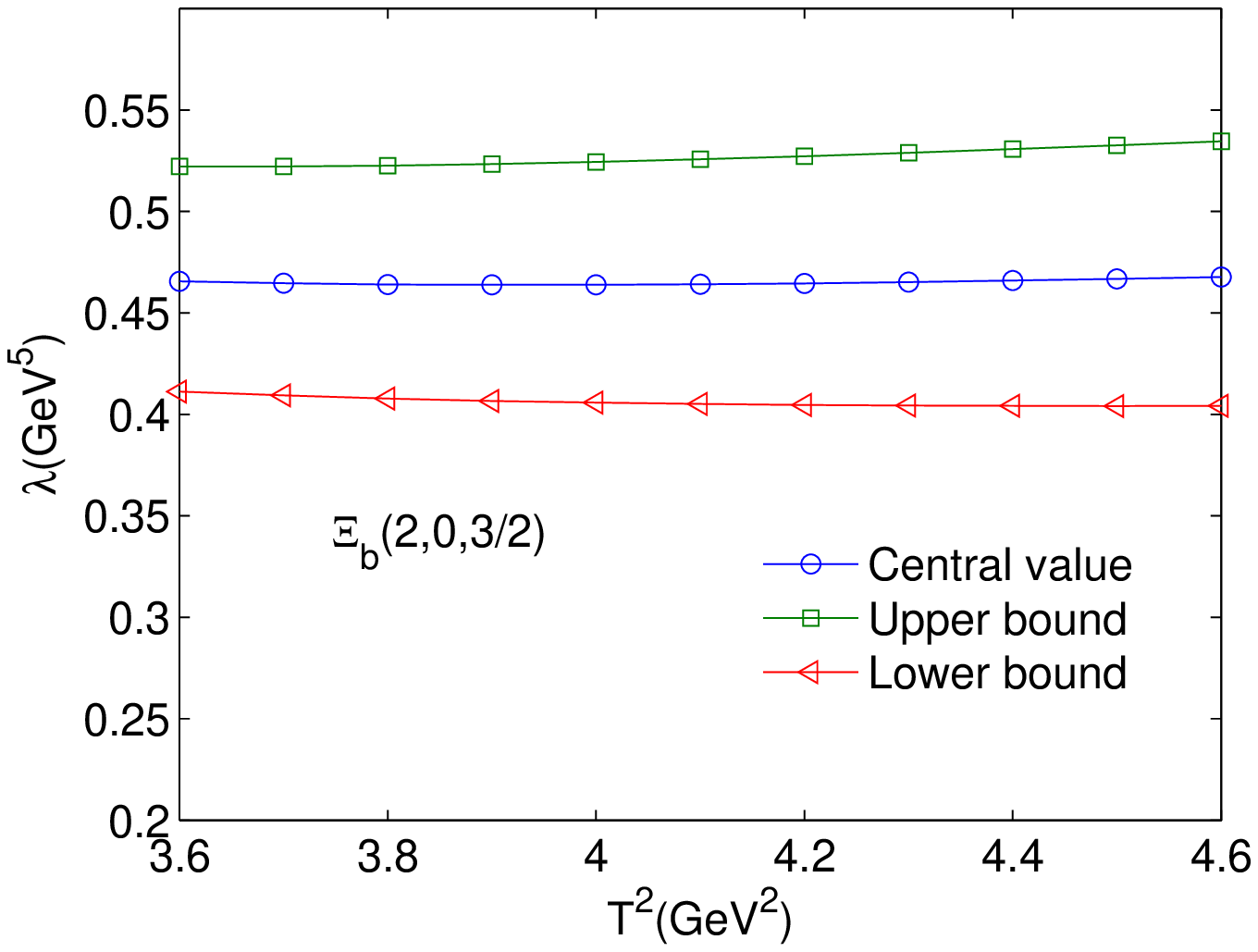}
\caption{The pole residues of the bottom baryon state $\Xi_{b}$(2,0,$\frac{3}{2}$) with variations of the Borel
parameters $T^{2}$}
\end{minipage}
\end{figure}
\begin{figure}[h]
\begin{minipage}[t]{0.45\linewidth}
\centering
\includegraphics[height=5cm,width=7cm]{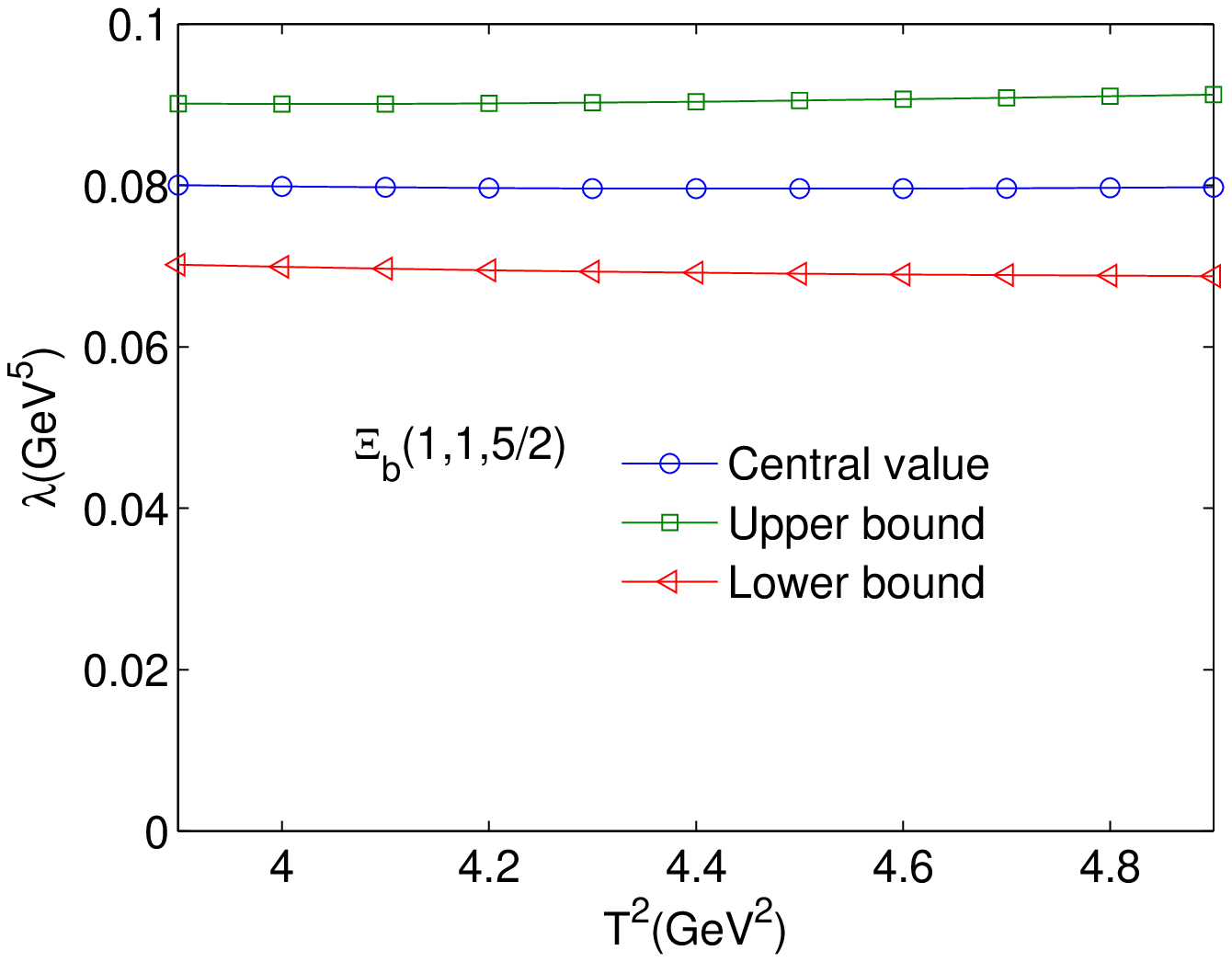}
\caption{The pole residues of the bottom baryon state $\Xi_{b}$(1,1,$\frac{5}{2}$) with variations of the Borel
parameters $T^{2}$}
\end{minipage}
\hfill
\begin{minipage}[t]{0.45\linewidth}
\centering
\includegraphics[height=5cm,width=7cm]{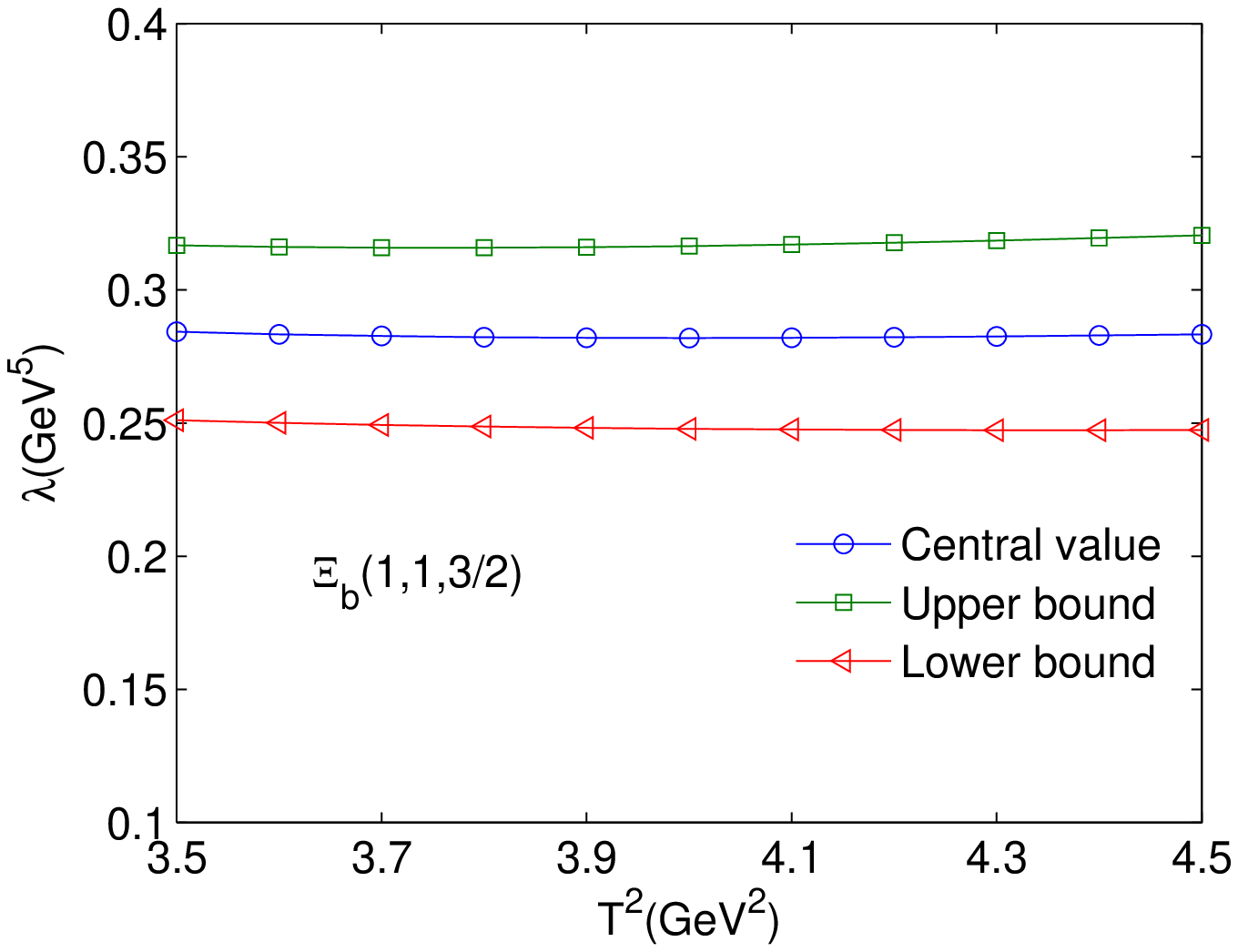}
\caption{The pole residues of the bottom baryon state $\Xi_{b}$(1,1,$\frac{3}{2}$) with variations of the Borel
parameters $T^{2}$}
\end{minipage}
\end{figure}
\begin{figure}[h]
\begin{minipage}[t]{0.45\linewidth}
\centering
\includegraphics[height=5cm,width=7cm]{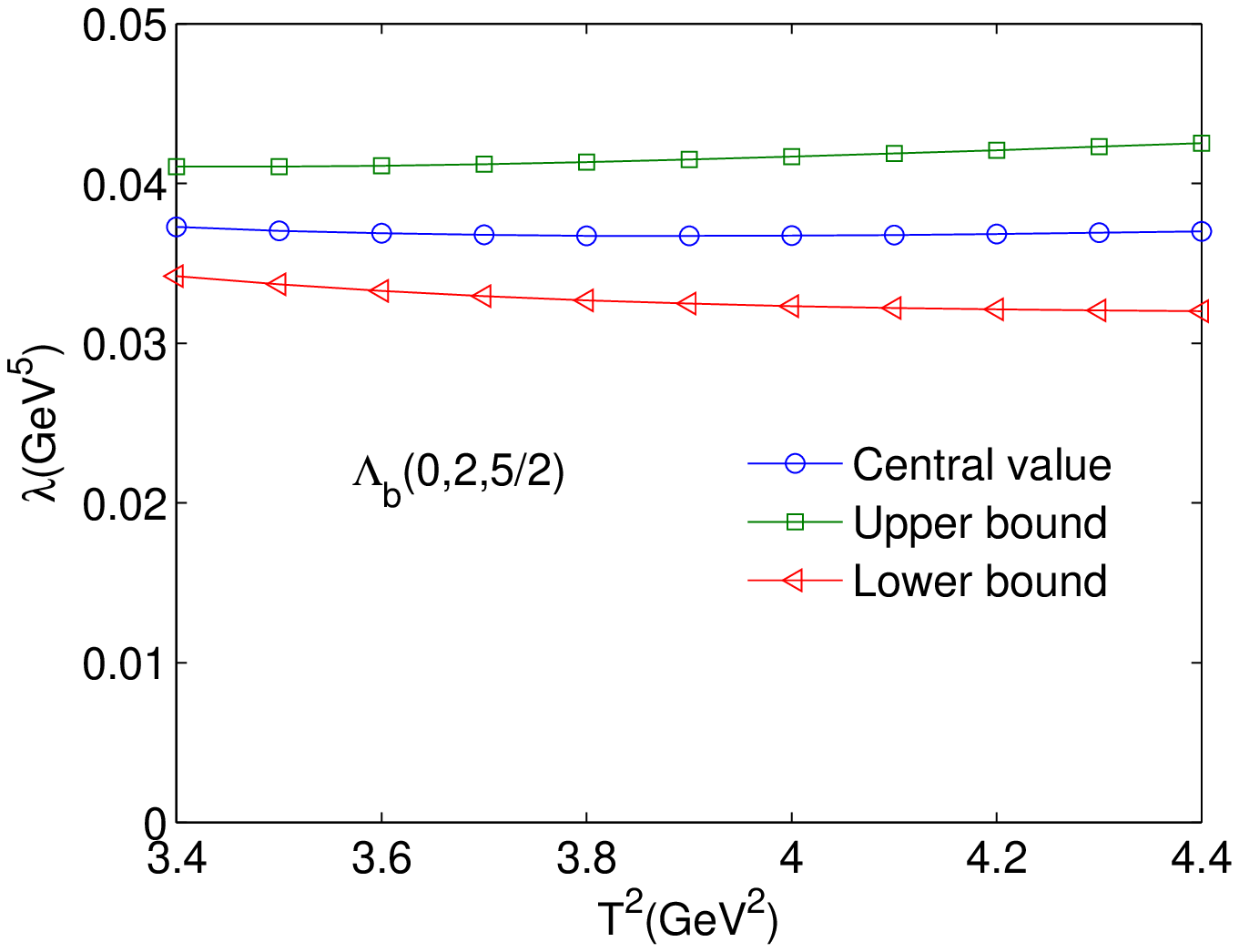}
\caption{The pole residues of the bottom baryon state $\Lambda_{b}$(0,2,$\frac{5}{2}$) with variations of the Borel
parameters $T^{2}$}
\end{minipage}
\hfill
\begin{minipage}[t]{0.45\linewidth}
\centering
\includegraphics[height=5cm,width=7cm]{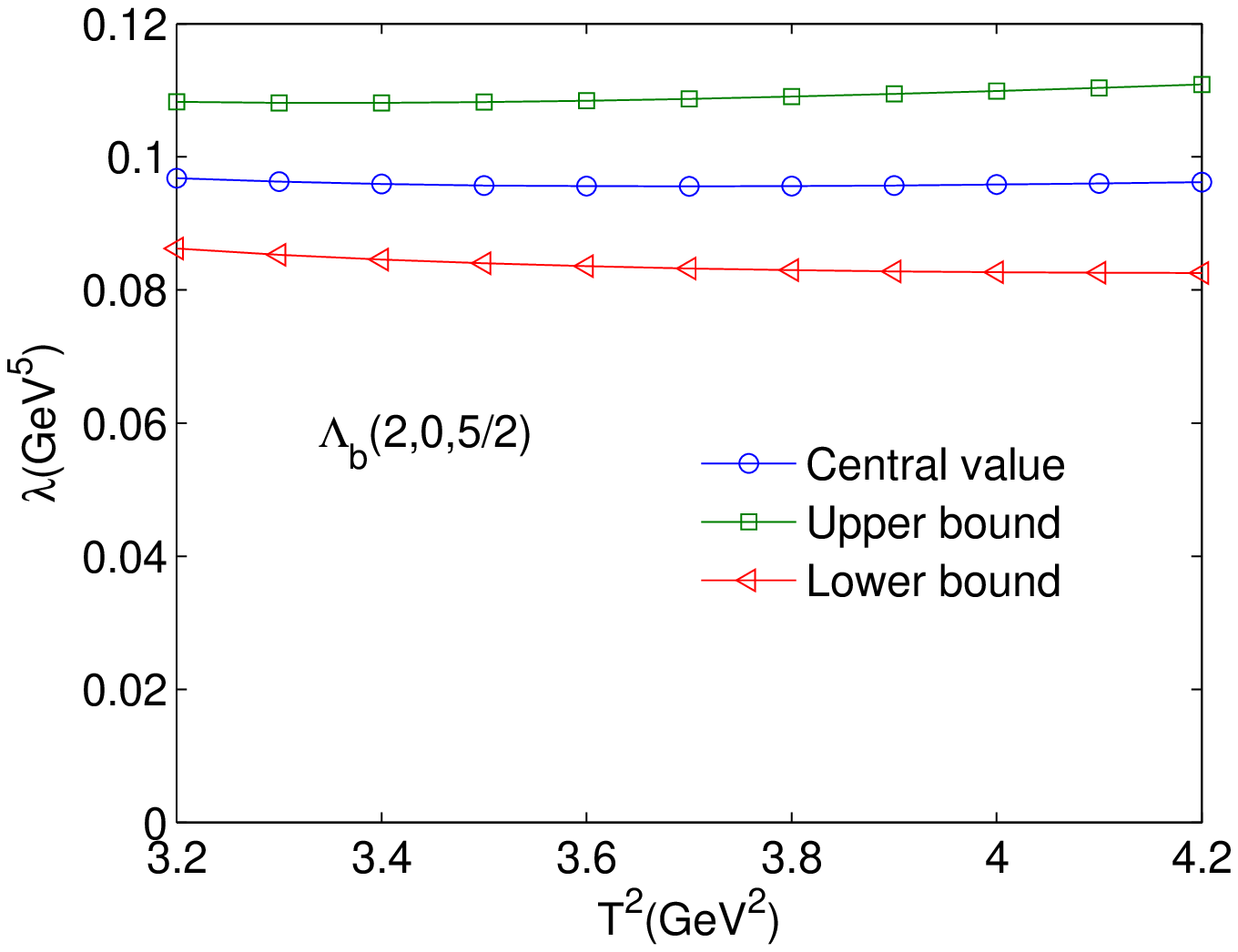}
\caption{The pole residues of the bottom baryon state $\Lambda_{b}$(2,0,$\frac{5}{2}$) with variations of the Borel
parameters $T^{2}$}
\end{minipage}
\end{figure}
\begin{figure}[h]
\begin{minipage}[t]{0.45\linewidth}
\centering
\includegraphics[height=5cm,width=7cm]{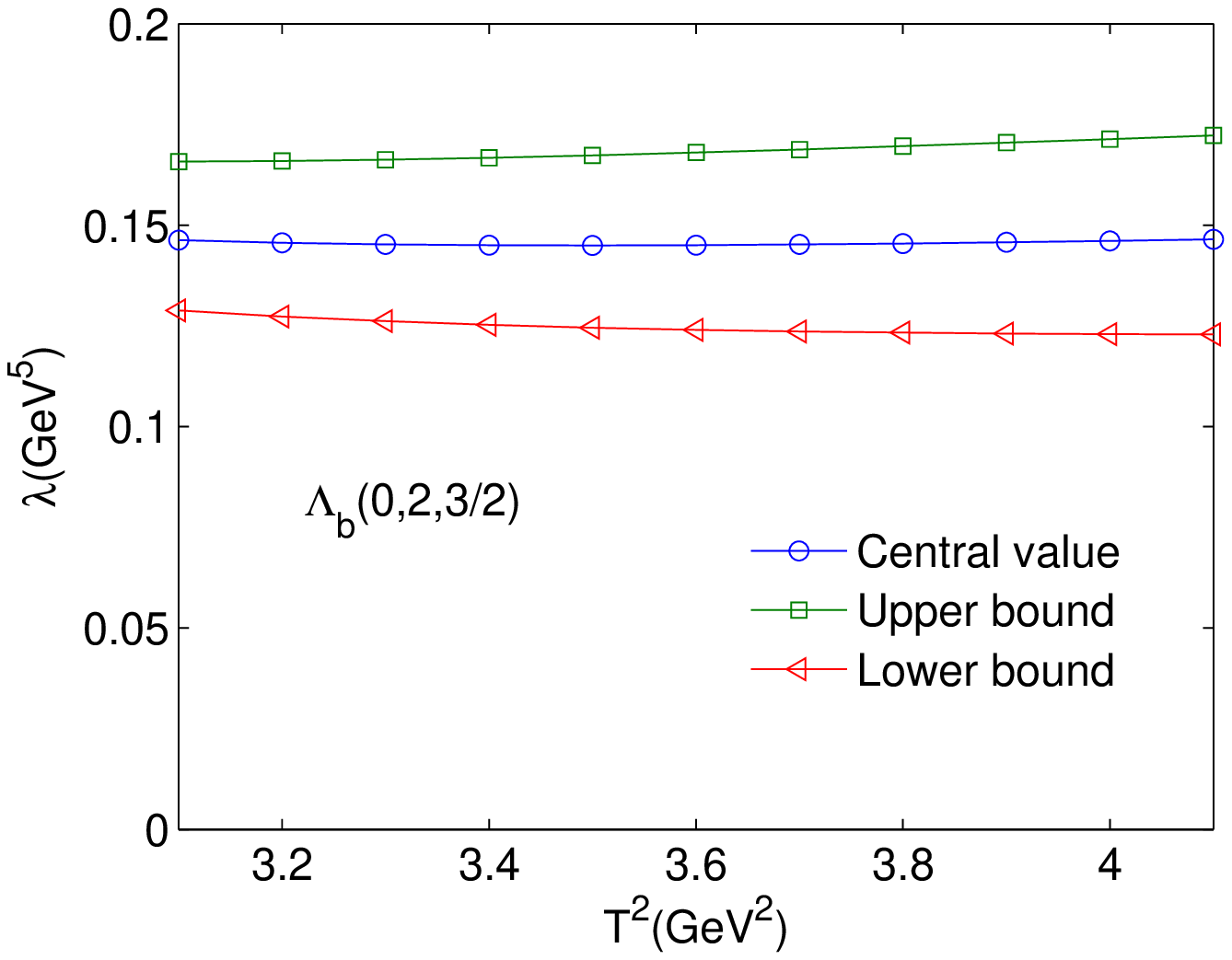}
\caption{The pole residues of the bottom baryon state $\Lambda_{b}$(0,2,$\frac{3}{2}$) with variations of the Borel
parameters $T^{2}$}
\end{minipage}
\hfill
\begin{minipage}[t]{0.45\linewidth}
\centering
\includegraphics[height=5cm,width=7cm]{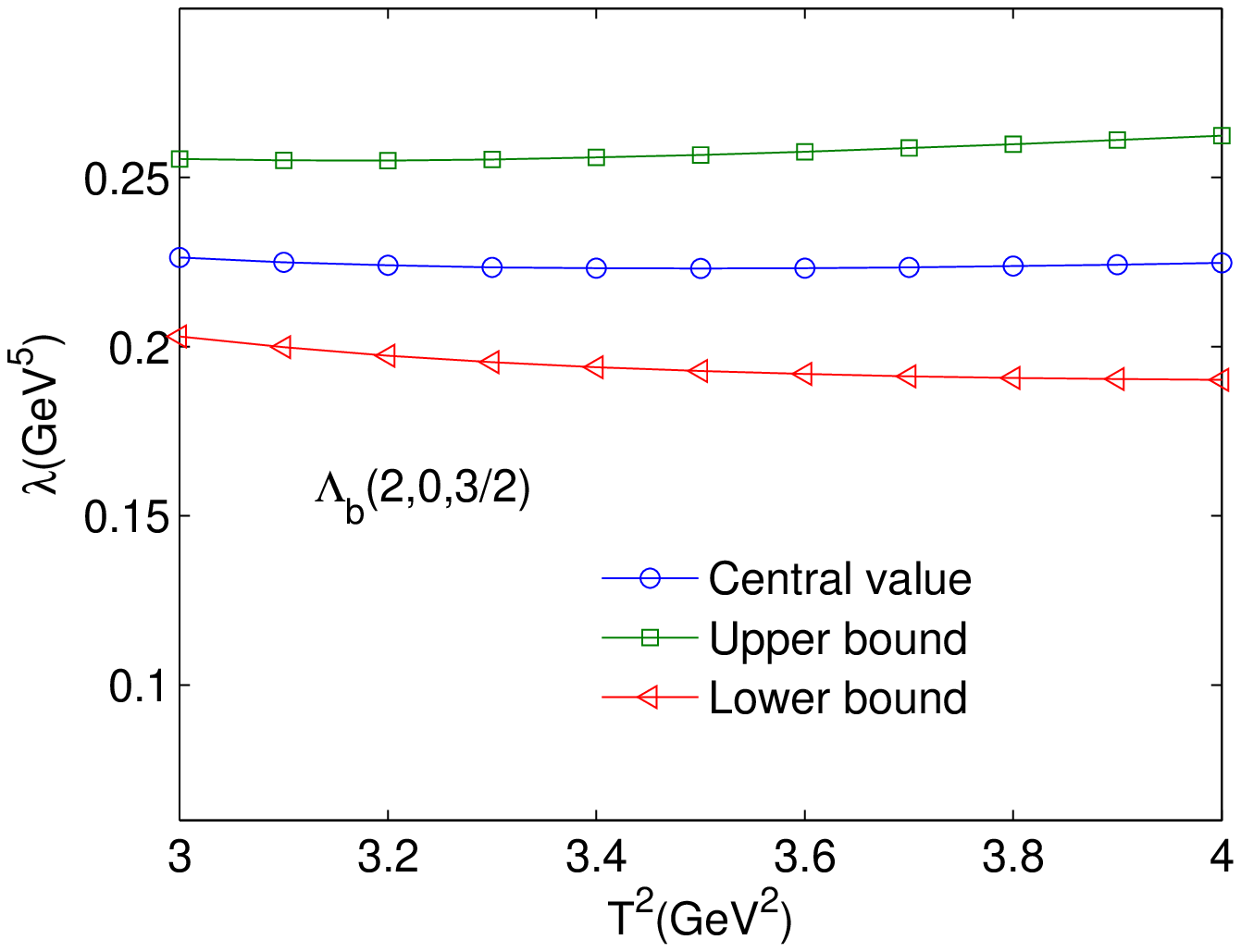}
\caption{The pole residues of the bottom baryon state $\Lambda_{b}$(2,0,$\frac{3}{2}$) with variations of the Borel
parameters $T^{2}$}
\end{minipage}
\end{figure}
\begin{figure}[h]
\begin{minipage}[t]{0.45\linewidth}
\centering
\includegraphics[height=5cm,width=7cm]{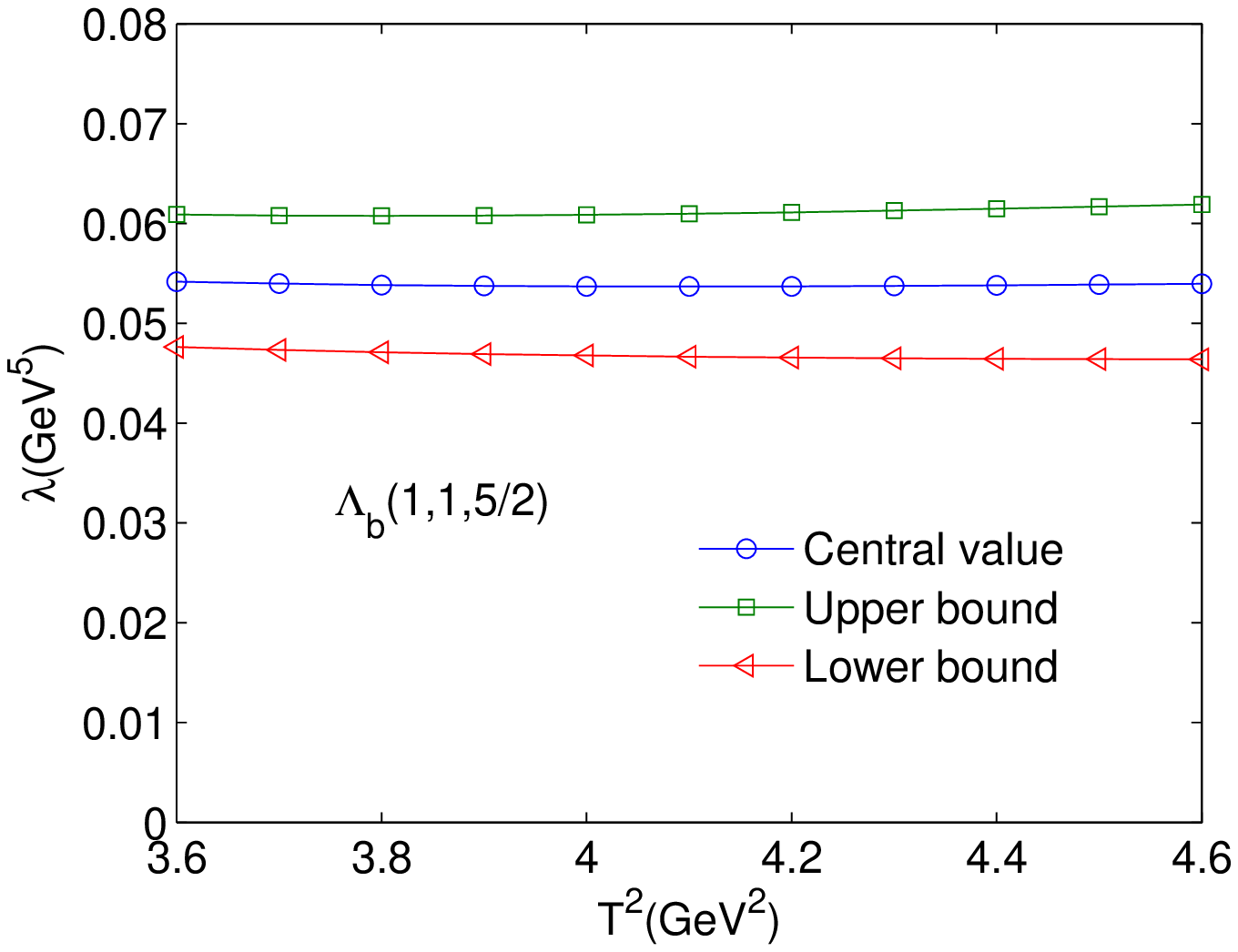}
\caption{The pole residues of the bottom baryon state $\Lambda_{b}$(1,1,$\frac{5}{2}$) with variations of the Borel
parameters $T^{2}$}
\end{minipage}
\hfill
\begin{minipage}[t]{0.45\linewidth}
\centering
\includegraphics[height=5cm,width=7cm]{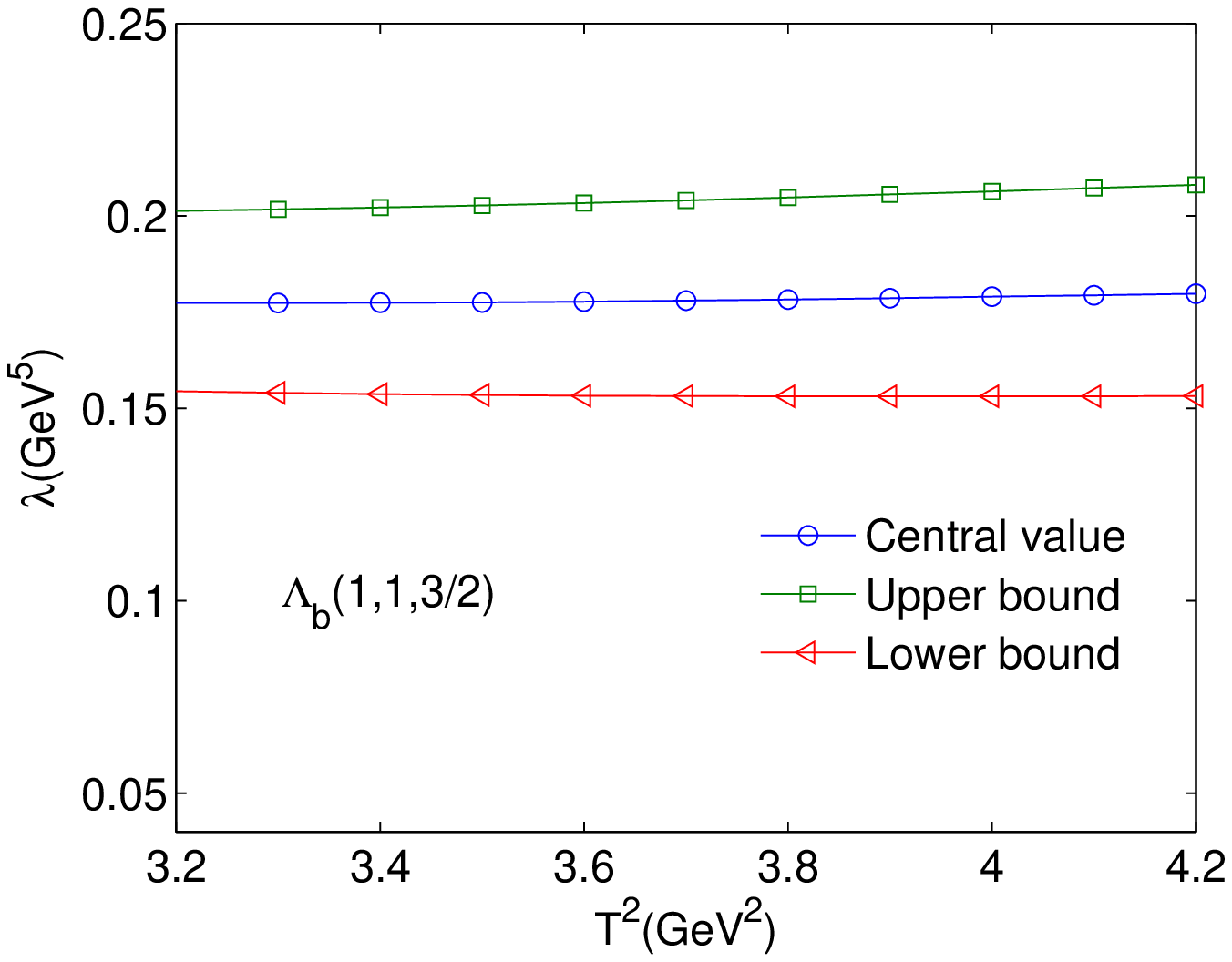}
\caption{The pole residues of the bottom baryon state $\Lambda_{b}$(1,1,$\frac{3}{2}$) with variations of the Borel
parameters $T^{2}$}
\end{minipage}
\end{figure}

The LHCb collaboration observed two structures with the masses of $m_{\Lambda_{b}}(6146)^{0}=6146.17\pm0.33\pm0.22\pm0.16$ MeV and
$m_{\Lambda_{b}}(6152)^{0}=6152.51\pm0.26\pm0.22\pm0.16$ MeV, and suggested their possible interpretation as a doublet of $\Lambda_{b}(1D)$ state. The quark-model predictions from different collaborations for the masses of this doublet ($\frac{3}{2}^{+}$,$\frac{5}{2}^{+}$) were ($6.145$ GeV, $6.165$ GeV)\cite{quam3}, ($6.190$ GeV, $6.196$ GeV)\cite{quam6}, ($6.181$ GeV, $6.183$ GeV)\cite{quam9} and ($6.147$ GeV, $6.153$ GeV)\cite{Theo2}. Our predictions for this doublet with the excitation mode ($L_{\rho},L_{\lambda}$)=($0,2$) are $m_{\Lambda_{b}}^{\frac{3}{2}^{+}}=6.13^{+0.10}_{-0.09}$ GeV  and $m_{\Lambda_{b}}^{\frac{5}{2}^{+}}=6.15^{+0.13}_{-0.15}$ GeV, respectively. This result is consistent with the experimental data\cite{LHCb2} and  quark-model predictions\cite{Theo2,quam3}, which supports assigning the $\Lambda_{b}(6146)$ and $\Lambda_{b}(6152)$ as the $1D$ $\Lambda_{b}$ doublet with the quantum numbers ($L_{\rho},L_{\lambda}$)=($0,2$) and $J^{p}=$$\frac{3}{2}^{+}$,$\frac{5}{2}^{+}$.

Up to now, the 1S, 1P, and 1D $\Lambda_{b}$ baryons have been established, but as for the $\Xi_{b}$ sector, only the ground state $\Xi_{b}(5797)$ has been confirmed\cite{Xi5797}. Especially, for radially excited $\Xi_{b}$ and $\Lambda_{b}$ states, fewer experimental results have been reported\cite{60720}. In Ref.\cite{quam6}, the mass spectra of $\Xi_{b}$ baryons were calculated in the  heavy-quark-light-diquark picture in the frame work of the QCD-motivated relativistic quark model. In Refs.\cite{Xi62271,L61460}, the masses and strong decay properties of $1D$ $\Xi_{b}$ baryons with  $J^{p}=$$\frac{3}{2}^{+}$ and $\frac{5}{2}^{+}$ were studied with the  quark model and $^{3}P_{0}$ model. Actually, these calculations with quark model were carried out on the basis of treating bottom baryons as the excitation mode ($L_{\rho},L_{\lambda}$)=($0,2$). Their predicted masses for $1D$ $\Xi_{b}$ doublet were ($6366$ MeV, $6373$ MeV) in Ref.\cite{quam6} and ($6327$ MeV, $6330$ MeV) in Refs.\cite{Xi62271,L61460}, respectively. From Table I, one can see that the QCD sum rule predictions for the masses of this doublet with excitation mode ($L_{\rho},L_{\lambda}$)=($0,2$) are $m_{\frac{3}{2}^{+}}=6.34^{+0.12}_{-0.11}$ GeV and $m_{\frac{5}{2}^{+}}=6.36^{+0.11}_{-0.12}$ GeV, which is consistent with the experiments\cite{Xi6333} and the predictions in Refs.\cite{Xi62271,L61460}. Thus it is reasonable to describe the $\Xi_{b}(6327)$ and $\Xi_{b}(6333)$ baryons as the $1D$($\Xi_{b}$) doublet with the excited mode ($L_{\rho},L_{\lambda}$)=($0,2$) and quantum numbers $J^{p}=$$\frac{3}{2}^{+}$ and $\frac{5}{2}^{+}$. For the $2D$ $\Lambda_{b}$ and $\Xi_{b}$ doublets, their masses with $\lambda-$mode were predicted as ($6526$ MeV, $6531$ MeV) and ($6690$ MeV, $6696$ MeV) in Ref.\cite{quam6}, which is roughly compatible with our results ($6.47^{+0.09}_{-0.10}$ GeV, $6.53^{+0.14}_{-0.14}$ GeV) and ($6.62^{+0.10}_{-0.13}$ GeV, $6.69^{+0.13}_{-0.11}$ GeV). From Tables I-II, we also notice that whether for $1D$ or $2D$ states, the prediction for the mass of the orbital excitation mode ($L_{\rho},L_{\lambda}$)=($0,2$) is little lower than those of the other excitation modes. Except the $1D$ $\Xi_{b}$ states, the predicted mass for the excitation mode ($L_{\rho},L_{\lambda}$)=($1,1$) is little higher than the others.

Finally, we would like to note that not only masses but also decay and production properties are useful to reveal the inner structrue of the heavy baryons. The predicted pole residues for the D-wave $\Xi_{b}$ and $\Lambda_{b}$ baryons in this paper are useful as an important parameter in studying the strong decay properties in the future. With the running of LHCb, we may expect these excited $\Xi_{b}$ and $\Lambda_{b}$ baryons to be observed in the near future.

\begin{large}
\textbf{4 Conclusions}
\end{large}

In summary, theoretical and experimental physicists have made great progresses in the field of single bottom baryons, such as the $\Lambda_{b}(6072)$\cite{60720,60721}, $\Lambda_{b}(6146)$\cite{L61460,L61461,L61462,L61463}, $\Lambda_{b}(6152)$\cite{L61460,L61461,L61462,L61463}, $\Xi_{b}(6227)$\cite{Xi62270,Xi62271,Xi62272}, $\Xi_{b}(6100)$\cite{Xi61000}, $\Xi_{b}(6327)$\cite{L61460,Xi6333} and $\Xi_{b}(6333)$\cite{L61460,Xi6333}. Stimulated by the observations of these new bottom states, we carry out a systematical study about the $1D$ and $2D$ $\Lambda_{b}$ and $\Xi_{b}$ baryons with the method of QCD sum rules. According to the heavy quark effective theory, we categorize the D-wave bottom baryons into three types which are denoted by their orbital excitation modes $(L_{\rho},L_{\lambda})=(0,2)$, $(2,0)$ and $(1,1)$. According to these excitation modes, we construct three-types interpolating currents to study the $1D$ as well as $2D$ bottom baryons with spin-parity $J^{p}=\frac{3}{2}^{+}$ and $\frac{5}{2}^{+}$. In our calculations, we successfully separate the contributions of the positive and negative states, which makes the QCD sum rules refrain from the contaminations of the bottom baryon states with negative parity. We carry out the operator product expansion(OPE) up to the vacuum condensates of dimension $10$ to warrant the reliability of the final results. Our predictions favor assigning the $\Lambda_{b}(6146)$ and $\Lambda_{b}(6152)$ as a $1D$ $\Lambda_{b}$ doublet with quantum numbers of $(L_{\rho},L_{\lambda})=(0,2)$ and $J^{P}=$($\frac{3}{2}^{+}$, $\frac{5}{2}^{+}$), respectively. This conclusion is consistent with experiments and with those of other collaborations\cite{LHCb2,Theo2,quam3}. As for the $\Xi_{b}$($1D$) states, we predict the masses of the excitation mode $(L_{\rho},L_{\lambda})=(0,2)$ as $m_{\frac{3}{2}^{+}}=6.34^{+0.12}_{-0.11}$ GeV and $m_{\frac{5}{2}^{+}}=6.36^{+0.11}_{-0.12}$ GeV. This result is compatible with the experimental data\cite{Xi6333} as well as the quark-model predictions\cite{Xi62271,L61460}. Thus, these two states can be interpreted as the $\Xi_{b}$($1D$) doublet with the quantum numbers $(L_{\rho},L_{\lambda})=(0,2)$ and $J^{P}=$$\frac{3}{2}^{+}$, $\frac{5}{2}^{+}$, respectively. Finally, our results show that the prediction for the mass of the excitation mode $(L_{\rho},L_{\lambda})=(0,2)$ is the smallest in these three excitation modes and the mass of $(L_{\rho},L_{\lambda})=(1,1)$ is the largest except $1D$ $\Xi_{b}$ state. As for the pole residues predicted in this paper, they are useful as an important parameter in studying the strong decay properties of the $1D$ and $2D$ $\Xi_{b}$ and $\Lambda_{b}$ states.

\begin{large}
\textbf{Acknowledgment}
\end{large}

This work has been supported by the Fundamental Research Funds for the Central Universities, Grant Number $2016MS133$, Natural Science Foundation of HeBei Province, Grant Number $A2018502124$.

\begin{center}
\begin{Large}
\textbf{Appendix}
\end{Large}
\end{center}
\begin{eqnarray}
\notag
\rho^{0,\Xi_{b}}_{\frac{5}{2},2,0}(s)=&&\int^{1}_{\frac{m^{2}_{b}}{s}}\Big\{\frac{1}{69120\pi^{4}}\big(132 - 494 x + 665 x^2 - 360 x^3 + 50 x^4 - 2 x^5 + 9 x^6\big)\times\big(s-\frac{m_{b}^{2}}{x}\big)^{4}\\
\notag
&&+\frac{m_{s}(5\langle \overline{s}s\rangle-2\langle \overline{q}q\rangle)}{96\pi^{2}}\times\big(x^2 - 2 x^3 + x^4\big)\times\big(s-\frac{m_{b}^{2}}{x}\big)^{2} \\
&& \notag
-\frac{m_{s}\langle \overline{q}g_{s}\sigma Gq\rangle}{36\pi^{2}}\times\big(4 - 7 x +3 x^2\big)\times\big(s-\frac{m_{b}^{2}}{x}\big)
\\
&& \notag
+\frac{m_{s}\langle \overline{s}g_{s}\sigma Gs\rangle}{216\pi^{2}}\times\big(31 - 58 x + 27 x^2\big)\times\big(s-\frac{m_{b}^{2}}{x}\big)
\\
&& \notag
+\frac{1}{34560\pi^{2}}\langle\frac{\alpha_{s}GG}{\pi}\rangle\times\big(665 + 132/x^2 - 494/x - 360 x + 50 x^2 - 2 x^3 + 9 x^4\big)\times\big(s-\frac{m_{b}^{2}}{x}\big)^{2}
\\
&& \notag
+\frac{1}{6192\pi^{2}}\langle\frac{\alpha_{s}GG}{\pi}\rangle\times\big(46 - 72 x + 15 x^2 + 2 x^3 + 9 x^4\big)\times\big(s-\frac{m_{b}^{2}}{x}\big)^{2}\Big\}dx
\\
&& +\frac{\langle \overline{q}g_{s}\sigma Gq\rangle\langle \overline{s}g_{s}\sigma Gs\rangle}{72}\delta\big(s-m_{b}^{2}\big)
\end{eqnarray}
\begin{eqnarray}
\notag
\rho^{1,\Xi_{b}}_{\frac{5}{2},2,0}(s)=&&\int^{1}_{\frac{m^{2}_{b}}{s}}\Big\{\frac{1}{69120\pi^{4}}\big(128 x - 457 x^2 + 575 x^3 - 290 x^4 + 70 x^5 - 53 x^6 + 27 x^7\big)\times\big(s-\frac{m_{b}^{2}}{x}\big)^{4}\\
\notag
&&-\frac{m_{s}(5\langle \overline{s}s\rangle-2\langle \overline{q}q\rangle)}{288\pi^{2}}\times\big(x^2 - 11 x^3 + 19 x^4 - 9 x^5\big)\times\big(s-\frac{m_{b}^{2}}{x}\big)^{2} \\
&& \notag
+\frac{m_{s}\langle \overline{q}g_{s}\sigma Gq\rangle}{72\pi^{2}}\times\big(x - 18 x^2 + 35 x^3 - 18 x^4\big)\times\big(s-\frac{m_{b}^{2}}{x}\big)
\\
&& \notag
-\frac{m_{s}\langle \overline{s}g_{s}\sigma Gs\rangle}{216\pi^{2}}\times\big(4 x - 74 x^2 + 151 x^3 - 81 x^4\big)\times\big(s-\frac{m_{b}^{2}}{x}\big)
\\
&& \notag
+\frac{1}{6192\pi^{2}}\langle\frac{\alpha_{s}GG}{\pi}\rangle\times\big(40 x - 57 x^2 + 21 x^3 - 31 x^4 + 27 x^5\big)\times\big(s-\frac{m_{b}^{2}}{x}\big)^{2}
\\
&& \notag
-\frac{m_{b}^{2}}{51840\pi^{2}}\langle\frac{\alpha_{s}GG}{\pi}\rangle\times\big(575 + 128/x^2 - 457/x - 290 x + 70 x^2 - 53 x^3 + 27 x^4\big)\times\big(s-\frac{m_{b}^{2}}{x}\big)\Big\}dx
\\
&& +\frac{5\langle \overline{q}g_{s}\sigma Gq\rangle\langle \overline{s}g_{s}\sigma Gs\rangle}{432}\delta\big(s-m_{b}^{2}\big)
\end{eqnarray}
\begin{eqnarray}
\notag
\rho^{0,\Xi_{b}}_{\frac{5}{2},0,2}(s)=&&\int^{1}_{\frac{m^{2}_{b}}{s}}\Big\{-\frac{1}{4608\pi^{4}}\big(2 x - 11 x^2 + 24 x^3 - 26 x^4 + 14 x^5 - 3 x^6\big)\times\big(s-\frac{m_{b}^{2}}{x}\big)^{4}\\
\notag
&&+\frac{m_{s}(\langle \overline{s}s\rangle-2\langle \overline{q}q\rangle)}{96\pi^{2}}\times\big(x^2 - 2 x^3 + x^4\big)\times\big(s-\frac{m_{b}^{2}}{x}\big)^{2} \\
&& \notag
+\frac{m_{b}^{2}}{3456\pi^{2}}\langle\frac{\alpha_{s}GG}{\pi}\rangle\times\big(24 + 2/x^2 - 11/x - 26 x + 14 x^2 - 3 x^3\big)\times\big(s-\frac{m_{b}^{2}}{x}\big)
\\
&& \notag
+\frac{1}{2304\pi^{2}}\langle\frac{\alpha_{s}GG}{\pi}\rangle\times\big(11 - 2/x - 24 x + 26 x^2 - 14 x^3 + 3 x^4\big)\times\big(s-\frac{m_{b}^{2}}{x}\big)^{2} \\
&& \notag
+\frac{1}{768\pi^{2}}\langle\frac{\alpha_{s}GG}{\pi}\rangle\times\big(x^2 - 2 x^3 + x^4\big)\times\big(s-\frac{m_{b}^{2}}{x}\big)^{2}
\Big\}dx
\\
&& +\frac{\langle \overline{q}g_{s}\sigma Gq\rangle\langle \overline{s}g_{s}\sigma Gs\rangle}{72}\delta\big(s-m_{b}^{2}\big)
\end{eqnarray}
\begin{eqnarray}
\notag
\rho^{1,\Xi_{b}}_{\frac{5}{2},0,2}(s)=&&\int^{1}_{\frac{m^{2}_{b}}{s}}\Big\{\frac{1}{4608\pi^{4}}\big(x^2 + 5 x^3 - 30 x^4 + 50 x^5 - 35 x^6 + 9 x^7\big)\times\big(s-\frac{m_{b}^{2}}{x}\big)^{4}\\
\notag
&&-\frac{m_{s}(\langle \overline{s}s\rangle-2\langle \overline{q}q\rangle)}{288\pi^{2}}\times\big(x^2 - 11 x^3 + 19 x^4 - 9 x^5\big)\times\big(s-\frac{m_{b}^{2}}{x}\big)^{2} \\
&& \notag
-\frac{m_{b}^{2}}{3456\pi^{2}}\langle\frac{\alpha_{s}GG}{\pi}\rangle\times\big(5 + 1/x - 30 x + 50 x^2 - 35 x^3 + 9 x^4\big)\times\big(s-\frac{m_{b}^{2}}{x}\big)
\\
&& \notag
-\frac{1}{2304\pi^{2}}\langle\frac{\alpha_{s}GG}{\pi}\rangle\times\big(x^2 - 11 x^3 + 19 x^4 - 9 x^5\big)\times\big(s-\frac{m_{b}^{2}}{x}\big)^{2}
\Big\}dx
\\
&& +\frac{5\langle \overline{q}g_{s}\sigma Gq\rangle\langle \overline{s}g_{s}\sigma Gs\rangle}{432}\delta\big(s-m_{b}^{2}\big)
\end{eqnarray}
\begin{eqnarray}
\notag
\rho^{0,\Xi_{b}}_{\frac{5}{2},1,1}(s)=&&\int^{1}_{\frac{m^{2}_{b}}{s}}\Big\{-\frac{1}{13824\pi^{4}}\big(3 - 20 x + 47 x^2 - 48 x^3 + 17 x^4 + 4 x^5 - 3 x^6\big)\times\big(s-\frac{m_{b}^{2}}{x}\big)^{4}\\
\notag
&&+\frac{m_{s}(3\langle \overline{s}s\rangle-2\langle \overline{q}q\rangle)}{96\pi^{2}}\times\big(x^2 - 2 x^3 + x^4\big)\times\big(s-\frac{m_{b}^{2}}{x}\big)^{2} \\
&& \notag
-\frac{m_{s}\langle \overline{q}g_{s}\sigma Gq\rangle}{48\pi^{2}}\times\big(x - 3 x^2 + 2 x^3\big)\times\big(s-\frac{m_{b}^{2}}{x}\big)
\\
&& \notag
+\frac{5m_{s}\langle \overline{s}g_{s}\sigma Gs\rangle}{432\pi^{2}}\times\big(x - 4 x^2 + 3 x^3\big)\times\big(s-\frac{m_{b}^{2}}{x}\big)
\\
&& \notag
-\frac{m_{b}^{2}}{10368\pi^{2}}\langle\frac{\alpha_{s}GG}{\pi}\rangle\times\big(48 - 3/x^3 + 20/x^2 - 47/x - 17 x - 4 x^2 + 3 x^3\big)\times\big(s-\frac{m_{b}^{2}}{x}\big)
\\
&& \notag
-\frac{1}{6192\pi^{2}}\langle\frac{\alpha_{s}GG}{\pi}\rangle\times\big(47 + 3/x^2 - 20/x - 48 x + 17 x^2 + 4 x^3 - 3 x^4\big)\times\big(s-\frac{m_{b}^{2}}{x}\big)^{2}
\\
&&
-\frac{1}{4608\pi^{2}}\langle\frac{\alpha_{s}GG}{\pi}\rangle\times\big(1 - 12 x + 15 x^2 + 2 x^3 - 6 x^4\big)\times\big(s-\frac{m_{b}^{2}}{x}\big)^{2}\Big\}dx
\end{eqnarray}
\begin{eqnarray}
\notag
\rho^{1,\Xi_{b}}_{\frac{5}{2},1,1}(s)=&&\int^{1}_{\frac{m^{2}_{b}}{s}}\Big\{\frac{1}{13824\pi^{4}}\big(2 x + 6 x^2 - 35 x^3 + 40 x^4 - 22 x^6 + 9 x^7\big)\times\big(s-\frac{m_{b}^{2}}{x}\big)^{4}\\
\notag
&&-\frac{m_{s}(3\langle \overline{s}s\rangle-2\langle \overline{q}q\rangle)}{288\pi^{2}}\times\big(x^2 - 11 x^3 + 19 x^4 - 9 x^5\big)\times\big(s-\frac{m_{b}^{2}}{x}\big)^{2} \\
&& \notag
+\frac{m_{s}\langle \overline{q}g_{s}\sigma Gq\rangle}{288\pi^{2}}\times\big(x - 22 x^2 + 57 x^3 - 36 x^4\big)\times\big(s-\frac{m_{b}^{2}}{x}\big)
\\
&& \notag
-\frac{m_{s}\langle \overline{s}g_{s}\sigma Gs\rangle}{432\pi^{2}}\times\big(x - 24 x^2 + 68 x^3 - 45 x^4\big)\times\big(s-\frac{m_{b}^{2}}{x}\big)
\\
&& \notag
+\frac{m_{b}^{2}}{10368\pi^{2}}\langle\frac{\alpha_{s}GG}{\pi}\rangle\times\big(35 - 2/x^2 - 6/x - 40 x + 22 x^3 - 9 x^4\big)\times\big(s-\frac{m_{b}^{2}}{x}\big)
\\
&&
+\frac{1}{4608\pi^{2}}\langle\frac{\alpha_{s}GG}{\pi}\rangle\times\big(x^2 + 16 x^3 - 35 x^4 + 18 x^5\big)\times\big(s-\frac{m_{b}^{2}}{x}\big)^{2}\Big\}dx
\end{eqnarray}
\begin{eqnarray}
\notag
\rho^{0,\Xi_{b}}_{\frac{3}{2},2,0}(s)=&&\int^{1}_{\frac{m^{2}_{b}}{s}}\Big\{\frac{1}{3072\pi^{4}}\big(33 - 128 x + 182 x^2 - 108 x^3 + 17 x^4 + 4 x^5\big)\times\big(s-\frac{m_{b}^{2}}{x}\big)^{4}\\
&& \notag
+\frac{7m_{s}\langle \overline{q}g_{s}\sigma Gq\rangle}{24\pi^{2}}\times\big(x- x^2\big)\times\big(s-\frac{m_{b}^{2}}{x}\big)
\\
&& \notag
-\frac{6m_{s}\langle \overline{s}g_{s}\sigma Gs\rangle}{24\pi^{2}}\times\big(x- x^2\big)\times\big(s-\frac{m_{b}^{2}}{x}\big)
\\
&& \notag
+\frac{m_{b}^{2}}{2304\pi^{2}}\langle\frac{\alpha_{s}GG}{\pi}\rangle\times\big(108 - 33/x^3 + 128/x^2 - 182/x - 17 x - 4 x^2\big)\times\big(s-\frac{m_{b}^{2}}{x}\big)
\\
&& \notag
+\frac{1}{1536\pi^{2}}\langle\frac{\alpha_{s}GG}{\pi}\rangle\times\big(182 + 33/x^2 - 128/x - 108 x + 17 x^2 + 4 x^3\big)\times\big(s-\frac{m_{b}^{2}}{x}\big)^{2}\\
&& \notag
+\frac{1}{768\pi^{2}}\langle\frac{\alpha_{s}GG}{\pi}\rangle\times\big(31 - 54 x + 15 x^2 + 8 x^3\big)\times\big(s-\frac{m_{b}^{2}}{x}\big)^{2}\Big\}dx
\\
&& +\frac{3\langle \overline{q}g_{s}\sigma Gq\rangle\langle \overline{s}g_{s}\sigma Gs\rangle}{32}\delta\big(s-m_{b}^{2}\big)
\end{eqnarray}
\begin{eqnarray}
\notag
\rho^{1,\Xi_{b}}_{\frac{3}{2},2,0}(s)=&&\int^{1}_{\frac{m^{2}_{b}}{s}}\Big\{\frac{5}{3072\pi^{4}}\big(9 x - 32 x^2 + 38 x^3 - 12 x^4 - 7 x^5 + 4 x^6\big)\times\big(s-\frac{m_{b}^{2}}{x}\big)^{4}\\
\notag
&&+\frac{m_{s}(25\langle \overline{s}s\rangle-10\langle \overline{q}q\rangle)}{16\pi^{2}}\times\big(x^2 - 2 x^3 + x^4\big)\times\big(s-\frac{m_{b}^{2}}{x}\big)^{2}\\
&& \notag
-\frac{m_{s}\langle \overline{q}g_{s}\sigma Gq\rangle}{4\pi^{2}}\times\big(4 - 11 x + 7 x^2\big)\times\big(s-\frac{m_{b}^{2}}{x}\big)
\\
&& \notag
+\frac{m_{s}\langle \overline{s}g_{s}\sigma Gs\rangle}{24\pi^{2}}\times\big(33 - 97 x + 64 x^2\big)\times\big(s-\frac{m_{b}^{2}}{x}\big)
\\
&& \notag
-\frac{5m_{b}^{2}}{2304\pi^{2}}\langle\frac{\alpha_{s}GG}{\pi}\rangle\times\big(38 + 9/x^2 - 32/x - 12 x - 7 x^2 + 4 x^3\big)\times\big(s-\frac{m_{b}^{2}}{x}\big)
\\
&& \notag
+\frac{1}{768\pi^{2}}\langle\frac{\alpha_{s}GG}{\pi}\rangle\times\big(35 - 36 x - 33 x^2 + 34 x^3\big)\times\big(s-\frac{m_{b}^{2}}{x}\big)^{2}\Big\}dx
\\
&& +\frac{5\langle \overline{q}g_{s}\sigma Gq\rangle\langle \overline{s}g_{s}\sigma Gs\rangle}{96}\delta\big(s-m_{b}^{2}\big)
\end{eqnarray}
\begin{eqnarray}
\notag
\rho^{0,\Xi_{b}}_{\frac{3}{2},0,2}(s)=&&\int^{1}_{\frac{m^{2}_{b}}{s}}\Big\{\frac{1}{1024\pi^{4}}\big(3 - 16 x + 34 x^2 - 36 x^3 + 19 x^4 - 4 x^5\big)\times\big(s-\frac{m_{b}^{2}}{x}\big)^{4}
\\
&& \notag
+\frac{m_{b}^{2}}{768\pi^{2}}\langle\frac{\alpha_{s}GG}{\pi}\rangle\times\big(36 - 3/x^3 + 16/x^2 - 34/x - 19 x + 4 x^2\big)\times\big(s-\frac{m_{b}^{2}}{x}\big)
\\
&& \notag
+\frac{1}{512\pi^{2}}\langle\frac{\alpha_{s}GG}{\pi}\rangle\times\big(4 + 3/x^2 - 16/x - 36 x + 19 x^2 - 4 x^3\big)\times\big(s-\frac{m_{b}^{2}}{x}\big)^{2}\Big\}dx
\\
&& +\frac{3\langle \overline{q}g_{s}\sigma Gq\rangle\langle \overline{s}g_{s}\sigma Gs\rangle}{32}\delta\big(s-m_{b}^{2}\big)
\end{eqnarray}
\begin{eqnarray}
\notag
\rho^{1,\Xi_{b}}_{\frac{3}{2},0,2}(s)=&&\int^{1}_{\frac{m^{2}_{b}}{s}}\Big\{\frac{7}{1024\pi^{4}}\big(1 - 10 x^2 + 20 x^3 - 15 x^4 + 4 x^5\big)\times\big(s-\frac{m_{b}^{2}}{x}\big)^{4}\\
\notag
&&+\frac{m_{s}(5\langle \overline{s}s\rangle-10\langle \overline{q}q\rangle)}{16\pi^{2}}\times\big(x^2 - 2 x^3 + x^4\big)\times\big(s-\frac{m_{b}^{2}}{x}\big)^{2}
\\
&& \notag
+\frac{7m_{b}^{7}}{768\pi^{2}}\langle\frac{\alpha_{s}GG}{\pi}\rangle\times\big(10 - 1/x^2 - 20 x + 15 x^2 - 4 x^3\big)\times\big(s-\frac{m_{b}^{2}}{x}\big)
\\
&& \notag
+\frac{5}{128\pi^{2}}\langle\frac{\alpha_{s}GG}{\pi}\rangle\times\big(x^2 - 2 x^3 + x^4\big)\times\big(s-\frac{m_{b}^{2}}{x}\big)^{2}\Big\}dx
\\
&& +\frac{5\langle \overline{q}g_{s}\sigma Gq\rangle\langle \overline{s}g_{s}\sigma Gs\rangle}{96}\delta\big(s-m_{b}^{2}\big)
\end{eqnarray}
\begin{eqnarray}
\notag
\rho^{0,\Xi_{b}}_{\frac{3}{2},1,1}(s)=&&\int^{1}_{\frac{m^{2}_{b}}{s}}\Big\{-\frac{1}{768\pi^{4}}\big(3 - 13 x + 22 x^2 - 18 x^3 + 7 x^4 - x^5\big)\times\big(s-\frac{m_{b}^{2}}{x}\big)^{4}
\\
&& \notag
-\frac{m_{s}\langle \overline{s}g_{s}\sigma Gs\rangle}{48\pi^{2}}\times\big(x - x^2\big)\times\big(s-\frac{m_{b}^{2}}{x}\big)
\\
&& \notag
-\frac{m_{b}^{m_{b}^{2}}}{576\pi^{2}}\langle\frac{\alpha_{s}GG}{\pi}\rangle\times\big(18 - 3/x^3 + 13/x^2 - 22/x - 7 x + x^2\big)\times\big(s-\frac{m_{b}^{2}}{x}\big)
\\
&& \notag
-\frac{1}{384\pi^{2}}\langle\frac{\alpha_{s}GG}{\pi}\rangle\times\big(22 + 3/x^2 - 13/x - 18 x + 7 x^2 - x^3\big)\times\big(s-\frac{m_{b}^{2}}{x}\big)^{2}
\\
&&
-\frac{1}{1024\pi^{2}}\langle\frac{\alpha_{s}GG}{\pi}\rangle\times\big(5 - 18 x + 21 x^2 - 8 x^3\big)\times\big(s-\frac{m_{b}^{2}}{x}\big)^{2}\Big\}dx
\end{eqnarray}
\begin{eqnarray}
\notag
\rho^{1,\Xi_{b}}_{\frac{3}{2},1,1}(s)=&&\int^{1}_{\frac{m^{2}_{b}}{s}}\Big\{\frac{1}{768\pi^{4}}\big(7 - 20 x + 10 x^2 + 20 x^3 - 25 x^4 + 8 x^5\big)\times\big(s-\frac{m_{b}^{2}}{x}\big)^{4}\\
\notag
&&+\frac{m_{s}(15\langle \overline{s}s\rangle-10\langle \overline{q}q\rangle)}{16\pi^{2}}\times\big(x^2 - 2 x^3 +  x^4 \big)\times\big(s-\frac{m_{b}^{2}}{x}\big)^{2} \\
&& \notag
-\frac{5m_{s}\langle \overline{q}g_{s}\sigma Gq\rangle}{16\pi^{2}}\times x\big(1 - 4 x + 3 x^2\big)\times\big(s-\frac{m_{b}^{2}}{x}\big)
\\
&& \notag
+\frac{m_{s}\langle \overline{s}g_{s}\sigma Gs\rangle}{48\pi^{2}}\times x\big(11 - 49 x + 38 x^2\big)\times\big(s-\frac{m_{b}^{2}}{x}\big)
\\
&& \notag
-\frac{m_{b}^{2}}{576\pi^{2}}\langle\frac{\alpha_{s}GG}{\pi}\rangle\times\big(10 + 7/x^2 - 20/x + 20 x - 25 x^2 + 8 x^3\big)\times\big(s-\frac{m_{b}^{2}}{x}\big)
\\
&&
+\frac{1}{1024\pi^{2}}\langle\frac{\alpha_{s}GG}{\pi}\rangle\times x\big(15 + 14 x - 73 x^2 + 44 x^3\big)\times\big(s-\frac{m_{b}^{2}}{x}\big)^{2}\Big\}dx
\end{eqnarray}

\begin{eqnarray}
\notag
\rho^{0,\Lambda_{b}}_{j,l_{\rho},l_{\lambda}}(s)=\rho^{0,\Xi_{b}}_{j,l_{\rho},l_{\lambda}}(s)|_{m_{s}\rightarrow0,\langle \overline{s}s\rangle\rightarrow\langle \overline{q}q\rangle,\langle \overline{s}g_{s}\sigma Gs\rangle\rightarrow\langle \overline{q}g_{q}\sigma Gs\rangle} \\
\rho^{1,\Lambda_{b}}_{j,l_{\rho},l_{\lambda}}(s)=\rho^{1,\Xi_{b}}_{j,l_{\rho},l_{\lambda}}(s)|_{m_{s}\rightarrow0,\langle \overline{s}s\rangle\rightarrow\langle \overline{q}q\rangle,\langle \overline{s}g_{s}\sigma Gs\rangle\rightarrow\langle \overline{q}g_{q}\sigma Gs\rangle}
\end{eqnarray}
\begin{eqnarray}
\notag
\rho_{j,QCD}^{0}=m_{b}\rho^{0,\Xi_{b}(\Lambda_{b})}_{j,l_{\rho},l_{\lambda}}(s)\\
\rho_{j,QCD}^{1}=\rho^{1,\Xi_{b}(\Lambda_{b})}_{j,l_{\rho},l_{\lambda}}(s)
\end{eqnarray}


\begin{thebibliography}{99}

\bibitem{LHCb1} R. Aaij et al. (LHCb Collaboration), Phys. Rev. Lett. 109, 172003(2012).

\bibitem{CDF1} T. A. Aaltonen et al. (CDF Collaboration), Phys. Rev. D 88,071101(2013).

\bibitem{60720} R. Aaij et al. (LHCb Collaboration), arXiv:2002.05112v3 [hep-ex](2020)

\bibitem{60721} K. Azizi, Y.Sarac, H.Sundu, Phys. Rev. D 102, 034007(2020)

\bibitem{Xi62270} R. Aaij et al.(LHCb Collaboration), Phys. Rev. Lett 121, 072002(2018).

\bibitem{Xi62271} Bing Chen, Ke-Wei Wei, Xiang Liu, and Ailin Zhang, Phys. Rev. D 98, 031502(2018).

\bibitem{Xi62272} Kai Lei Wang, Qi Fang L\"{u}, and Xian Hui Zhong, Phys. Rev. D 99, 014011(2019).

\bibitem{Xi61000} A. M. Sirunyan et al.(CMS Collaboration), Phys. Rev. Lett 126, 252003(2021)

\bibitem{LHCb2} R. Aaij et al. (LHCb Collaboration), Phys. Rev. Lett. 123, 152001(2019).

\bibitem{L61460} Bing Chen, Si-Qiang Luo, Xiang Liu, and Takayuki Matsuki, Phys. Rev. D 100, 094032(2019)

\bibitem{L61461} H. M. Yang et al., arXiv:1909.13575(2019).

\bibitem{L61462} K. L. Wang, Q. F. L\"{u}, and X. H. Zhong, Phys. Rev. D 100, 114035(2019), arXiv:1908.04622.

\bibitem{L61463} W. Liang, Q. F. L\"{u}, and X. H. Zhong, Phys. Rev. D 100, 054013(2019), arXiv:1908.00223.

\bibitem{quam3} S. Capstick and N. Isgur, Phys. Rev. D 34, 2809 (1986); AIPConf. Proc. 132, 267(1985).

\bibitem{quam6} D. Ebert, R. N. Faustov, and V. O. Galkin, Phys. Rev. D 84, 014025(2011).

\bibitem{quam9} W. Roberts and M. Pervin, Int. J. Mod. Phys. A 23, 2817(2008).

\bibitem{Theo2} B. Chen, K.-W. Wei, and A. Zhang, Eur. Phys. J. A 51, 82(2015).

\bibitem{Xi6333}H. J. Mu et al.(LHCb Collaboration), Beauty-hadron spectroscopy at LHCb, EPS-HEP Conference 2021(2021).

\bibitem{quam1} L. A. Copley, N. Isgur, and G. Karl, Phys. Rev. D 20, 768(1979); 23, 817(E)(1981).

\bibitem{quam2} K. Maltman and N. Isgur, Phys. Rev. D 22, 1701(1980).

\bibitem{quam4} D. Ebert, R. N. Faustov, and V. O. Galkin, Phys. Rev. D 72, 034026(2005).

\bibitem{quam5} D. Ebert, R. N. Faustov, and V. O. Galkin, Phys. Lett. B 659, 612(2008).

\bibitem{quam7} H. Garcilazo, J. Vijande, and A. Valcarce, J. Phys. G 34, 961(2007).

\bibitem{quam8} A. Valcarce, H. Garcilazo, and J. Vijande, Eur. Phys. J. A37, 217(2008).

\bibitem{quam10} T. Yoshida, E. Hiyama, A. Hosaka, M. Oka, and K. Sadato,Phys. Rev. D 92, 114029(2015).

\bibitem{quam11} M. Karliner and J. L. Rosner, Phys. Rev. D 92, 074026(2015).

\bibitem{quam12} K. Thakkar, Z. Shah, A. K. Rai, and P. C. Vinodkumar,Nucl. Phys. A 965, 57(2017).

\bibitem{quam13} Z. Shah, K. Thakkar, A. K. Rai, and P. C. Vinodkumar,Chin. Phys. C 40, 123102(2016).

\bibitem{quam14} Z. Shah, K. Thakkar, A. Kumar Rai, and P. C. Vinodkumar,Eur. Phys. J. A 52, 313(2016).

\bibitem{quam15} F. Hussain, J. G. Korner, and S. Tawfiq, Phys. Rev. D 61, 114003(2000).

\bibitem{quam16} M. A. Ivanov, J. G. Korner, and V. E. Lyubovitskij, Phys.Lett. B 448, 143(1999).

\bibitem{quam17} M. A. Ivanov, J. G. Korner, V. E. Lyubovitskij,Rusetsky, Phys. Rev. D 60, 094002(1999).

\bibitem{quam18} C. Albertus, E. Hernandez, J. Nieves, and J. M. Verde-Velasco, Phys. Rev. D 72, 094022(2005).

\bibitem{quam19} S. Migura, D. Merten, B. Metsch, and H. Petry, Eur. Phys. J.A 28, 41(2006).

\bibitem{quam20} X. H. Zhong and Q. Zhao, Phys. Rev. D 77, 074008(2008).

\bibitem{quam21} E. Hernandez and J. Nieves, Phys. Rev. D 84, 057902(2011).

\bibitem{quam22} L. H. Liu, L. Y. Xiao, and X. H. Zhong, Phys. Rev. D 86, 034024(2012).

\bibitem{quam23} B. Chen, K. W. Wei, X. Liu, and T. Matsuki, Eur. Phys. J. C 77, 154(2017).

\bibitem{quam26} B. Chen and X. Liu, Phys. Rev. D 98, 074032(2018).

\bibitem{quam27} H. Nagahiro, S. Yasui, A. Hosaka, M. Oka, and H. Noumi, Phys. Rev. D 95, 014023(2017).

\bibitem{quam28} Y. X. Yao, K. L. Wang, and X. H. Zhong, Phys. Rev. D 98, 076015(2018).

\bibitem{chiral1} M. Q. Huang, Y. B. Dai, and C. S. Huang, Phys. Rev. D 52, 3986(1995); 55, 7317(E)(1997).

\bibitem{chiral2} M. C. Banuls, A. Pich, and I. Scimemi, Phys. Rev. D 61, 094009(2000).

\bibitem{chiral3} H. Y. Cheng and C. K. Chua, Phys. Rev. D 75, 014006(2007).

\bibitem{chiral4} N. Jiang, X. L. Chen, and S. L. Zhu, Phys. Rev. D 92, 054017(2015).

\bibitem{chiral5} H. Y. Cheng and C. K. Chua, Phys. Rev. D 92, 074014(2015).

\bibitem{chiral6} Y. Kawakami and M. Harada, Phys. Rev. D 99, 094016(2019).

\bibitem{3P01} C. Chen, X. L. Chen, X. Liu, W. Z. Deng, and S. L. Zhu, Phys. Rev. D 75, 094017(2007).

\bibitem{3P02} D. D. Ye, Z. Zhao, and A. Zhang, Phys. Rev. D 96, 114009(2017).

\bibitem{3P03} D. D. Ye, Z. Zhao, and A. Zhang, Phys. Rev. D 96, 114003(2017).

\bibitem{3P04} B. Chen, X. Liu, and A. Zhang, Phys. Rev. D 95, 074022(2017).

\bibitem{3P05} P. Yang, J. J. Guo, and A. Zhang, Phys. Rev. D 99, 034018(2019).

\bibitem{3P06} J. J. Guo, P. Yang, and A. Zhang, Phys. Rev. D 100, 014001(2019).

\bibitem{3P07} Q. F. L\"{u} and X. H. Zhong, Phys. Rev. D 101, 014017(2020).

\bibitem{Lattice1} M. Padmanath, R. G. Edwards, N. Mathur, and M. Peardon, arXiv:1311.4806(2013).

\bibitem{Lattice2} H. Bahtiyar, K. U. Can, G. Erkol, and M. Oka, Phys. Lett. B747, 281(2015).

\bibitem{Lattice3} P. Perez-Rubio, S. Collins, and G. S. Bali, Phys. Rev. D 92,034504(2015).

\bibitem{Lattice4} H. Bahtiyar, K. U. Can, G. Erkol, M. Oka, and T. T.Takahashi, Phys. Lett. B 772, 121(2017).

\bibitem{LCsum1} S. L. Zhu and Y. B. Dai, Phys. Rev. D 59, 114015(1999).

\bibitem{LCsum2} S. S. Agaev, K. Azizi, and H. Sundu, Phys. Rev. D 96, 094011(2017).

\bibitem{LCsum3} H. X. Chen, Q. Mao, W. Chen, A. Hosaka, X. Liu, and S. L.Zhu, Phys. Rev. D 95, 094008(2017).

\bibitem{LCsum4} Z. G. Wang, Phys. Rev. D 81, 036002(2010).

\bibitem{LCsum5} Z. G. Wang, Eur. Phys. J. A 44, 105(2010).

\bibitem{LCsum6} T. M. Aliev, K. Azizi, and H. Sundu, Eur. Phys. J. C 75, 14(2015).

\bibitem{LCsum7} T. M. Aliev, T. Barakat, and M. Savci, Phys. Rev. D 93, 056007(2016).

\bibitem{LCsum8} T. M. Aliev, K. Azizi, Y. Sarac, and H. Sundu, Phys. Rev. D 99, 094003(2019).

\bibitem{Sum1} S. L. Zhu, Phys. Rev. D 61, 114019(2000).

\bibitem{Sum2} Z. G. Wang, Eur. Phys. J. A 47, 81(2011).

\bibitem{Sum3} Q. Mao, H. X. Chen, W. Chen, A. Hosaka, X. Liu, and S. L.Zhu, Phys. Rev. D 92, 114007(2015).

\bibitem{Sum4} H. X. Chen, Q. Mao, A. Hosaka, X. Liu, and S. L. Zhu,Phys. Rev. D 94, 114016(2016).

\bibitem{Sum5} Z. G. Wang, Nucl. Phys. B 926, 467(2018).

\bibitem{Sum6} Q. Mao, H. X. Chen, A. Hosaka, X. Liu, and S. L. Zhu,Phys. Rev. D 96, 074021(2017).

\bibitem{Sum7} T. M. Aliev, K. Azizi, Y. Sarac, and H. Sundu, Phys. Rev. D98, 094014(2018).

\bibitem{Sum8} E. L. Cui, H. M. Yang, H. X. Chen, and A. Hosaka, Phys.Rev. D 99, 094021(2019).

\bibitem{Sum9} K. Azizi, Y. Sarac, and H. Sundu, Phys. Rev. D 101, 074026(2020).

\bibitem{Sum10} X. Liu, H. X. Chen, Y. R. Liu, A. Hosaka and S. L. Zhu, Phys. Rev. D 77, 014031(2008).

\bibitem{Theo1} E. E. Jenkins, Phys. Rev. D 77, 034012(2008).

\bibitem{Theo3} I. L. Grach, I. M. Narodetskii, M. A. Trusov, and A. I. Veselov, in Proceedings of the 18th International Conference on Particles and Nuclei (PANIC08)(2008)[arXiv:0811.2184].

\bibitem{Theo4} Z. Y. Wang, J. J. Qi, X. H. Guo, and K. W. Wei, Chin. Phys. C 41, 093103(2017).

\bibitem{Theo5} J. G. Korner, M. Kramer, and D. Pirjol, Prog. Part. Nucl.Phys. 33, 787(1994).

\bibitem{Theo6} J. M. Richard, Phys. Rep. 212, 1(1992).

\bibitem{Theo7} E. Klempt and J. M. Richard, Rev. Mod. Phys. 82, 1095(2010).

\bibitem{Theo8} H. X. Chen, W. Chen, X. Liu, Y. R. Liu, and S. L. Zhu, Rep.Prog. Phys. 80, 076201 (2017).

\bibitem{Theo9} H. Y. Cheng, Front. Phys. 10, 101406(2015).

\bibitem{Theo10} V. Crede and W. Roberts, Rep. Prog. Phys. 76, 076301(2013).

\bibitem{Xi5797} P. A. Zylaet al.(Particle Data Group), Prog. Theor. Exp. Phys.2020, 083C01 (2020) and 2021 update

\bibitem{sumrule1}B. L. Ioffe, Nucl. Phys. B 188, 317 (1981); B 191, 591(E)(1981).

\bibitem{sumrule2}V. M. Belyaev and B. L. Ioffe, Sov. Phys. JETP 56, 493(1982).

\bibitem{baryon1} Z. G. Wang, Commun. Theor. Phys. 58, 723(2012).

\bibitem{baryon2} K. Azizi and H. Sundu, Eur. Phys. J. Plus 132, 22(2017).

\bibitem{baryon3} Z. G. Wang, Phys. Lett. B 685, 59(2010).

\bibitem{baryon4}Z. G. Wang, Eur. Phys. J. C 68, 459(2010).

\bibitem{baryon5}Z. G. Wang, Eur. Phys. J. A 45, 267(2010); 47, (2011) 81

\bibitem{baryon6} H. X. Chen, W. Chen, Q. Mao, A. Hosaka, X. Liu and S. L. Zhu, Phys. Rev. D 91, 054034(2015).

\bibitem{gly}G. L. Yu, Z. G. Wang, INT J MOD PHYS A, 34(26), 1950151(2019)

\bibitem{sumrule3}R. Khosravi, M. Janbazi, Phys. Rev. D 87, 016003(2013);89, 016001(2014)

\bibitem{sumrule4}G. L. Yu, Z. G. Wang, Z. Y. Li, Chin. Phys. C 41(8), 083104(2017).

\bibitem{sumrule5} G. L. Yu, Z. G. Wang, Z. Y. Li, INT J MOD PHYS A 32(5), 1750203(2017).

\bibitem{sumrule6}X. Liu, Z. G. Luo, Z. F. Sun, Phys. Rev. Lett. 104, 122001(2010).

\bibitem{sumrule7}J. He, X. Liu, Phys. Rev. D 82, 114029(2010).

\bibitem{sumrule8} W. Chen, H. Y. Jin, R.T. Kleiv, et al, Phys. Rev. D 88, 045027(2013).

\bibitem{sumrule9}J. R. Zhang, M. Q. Huang, Phys. Lett. B 674, 28(2009).

\bibitem{sumrule10} J. R. Zhang, M. Q. Huang, Chin. Phys. C 33, 1385(2009).

\bibitem{WZG0}Z. G. Wang, Nucl. Phys. B 926,467(2018) arXiv:1705.07745v3 [hep-ph].

\bibitem{WZG1}Z. G. Wang, Eur. Phys. J. C 77,325(2017).

\bibitem{WZG2}Z. G. Wang, Eur. Phys. J. C 77,832(2017).

\bibitem{WZG3}Z. G. Wang, Int. J. Mod. Phys. A 35, 2050043(2020), arXiv:2001.02961v2 [hep-ph].

\bibitem{current}H. X. Chen, Q. Mao, A. Hosaka, X. Liu and S. L. Zhu, Phys. Rev. D 94, 114016 (2016).

\bibitem{parameters1} M. A. Shifman, A. I. Vainshtein and V. I. Zakharov, Nucl. Phys. B147 (1979) 385, 448.

\bibitem{parameters2}L. J. Reinders, H. Rubinstein and S. Yazaki, Phys. Rept. 127,1(1985).

\bibitem{energy1} Z. G. Wang and T. Huang, Phys. Rev. D 89,054019(2014.

\bibitem{energy2}Z. G. Wang, Eur. Phys. J. C 74, 2874(2014); 74, 2891(2014); 74, 2963(2014).

\bibitem{energy3}Z. G. Wang and T. Huang, Nucl. Phys. A 930, 63(2014).

\bibitem{energy4} Z. G. Wang, Eur. Phys. J. C 76, 70(2016).

\bibitem{energy5} Z. G. Wang, Eur. Phys. J. C, 75, 359(2015);77, 325(2017).

\bibitem{energy6} Z. G. Wang, Eur. Phys. J. C 79, 489 (2019).

\end{thebibliography}
\end{document}